\documentclass{jss}
\usepackage{thumbpdf,lmodern}
\usepackage{booktabs}
\usepackage{tabularx}
\usepackage{bm}
\usepackage{enumitem}
\IfFileExists{placeins.sty}{\usepackage{placeins}}{}
\graphicspath{{analysis/manuscript/outputs/figures/}}

\RequirePackage{amsthm,amsmath,amsfonts,amssymb}

\newcommand{\tworowcell}[1]{\smash{\raisebox{-0.5\normalbaselineskip}{#1}}}
\newcommand{\TableStyle}{\footnotesize\setlength{\tabcolsep}{4pt}\renewcommand{\arraystretch}{1.08}}
\newcommand{\CompactTableStyle}{\footnotesize\setlength{\tabcolsep}{2.3pt}\renewcommand{\arraystretch}{1.0}}
\newcommand{\vect}[1]{\boldsymbol{#1}}
\newcommand{\mat}[1]{\mathbf{#1}}
\newcommand{\N}{\mathcal{N}}
\newcommand{\IG}{\operatorname{IG}}
\newcommand{\Exp}{\operatorname{Exp}}
\newcommand{\AL}{\operatorname{AL}}
\newcommand{\exAL}{\operatorname{exAL}}
\newcommand{\KL}{\operatorname{KL}}
\newcommand{\CRPS}{\operatorname{CRPS}}
\newcommand{\PPLC}{\operatorname{PPLC}}
\newcommand{\R}{\mathbb{R}}

\newcommand{\Ind}{\mathbb{I}}

\renewcommand{\E}{\mathbb{E}}
\newcommand{\Var}{\operatorname{Var}}

\newcommand{\tr}{\operatorname{tr}}
\newcommand{\diag}{\operatorname{diag}}
\newcommand{\logit}{\operatorname{logit}}
\newcommand{\expit}{\operatorname{expit}}
\newcommand{\TNp}{\mathcal{N}^{+}}
\newcommand{\GIG}{\operatorname{GIG}}
\newcommand{\elbo}{\mathcal{L}}
\newcommand{\pigamma}{\pi_\gamma}
\newcounter{appalgorithm}
\renewcommand{\theappalgorithm}{\thesection.\arabic{appalgorithm}}
\newenvironment{appalgorithm}[1]{%
  \refstepcounter{appalgorithm}%
  \par\medskip\noindent\textbf{Algorithm~\theappalgorithm. #1.}\par\nobreak
}{\par\medskip}
\makeatletter
\newcommand{\exdqlmAppendixSetup}{%
  \setcounter{section}{0}%
  \setcounter{subsection}{0}%
  \setcounter{subsubsection}{0}%
  \setcounter{equation}{0}%
  \setcounter{table}{0}%
  \setcounter{appalgorithm}{0}%
  \@addtoreset{equation}{section}%
  \@addtoreset{table}{section}%
  \@addtoreset{appalgorithm}{section}%
  \renewcommand{\thesection}{\Alph{section}}%
  \renewcommand{\thesubsection}{\thesection.\arabic{subsection}}%
  \renewcommand{\thesubsubsection}{\thesubsection.\arabic{subsubsection}}%
  \renewcommand{\theequation}{\thesection.\arabic{equation}}%
  \renewcommand{\thetable}{\thesection.\arabic{table}}%
  \renewcommand{\theappalgorithm}{\thesection.\arabic{appalgorithm}}%
}
\makeatother

\author{Antonio De Leon \\ University of California Santa Cruz \And
        Raquel Barata \\ University of California Santa Cruz \AND
        Raquel Prado \\ University of California Santa Cruz \And
        Bruno Sans\'{o} \\ University of California Santa Cruz}
\title{\pkg{exdqlm}: An \proglang{R} Package for Estimation and Analysis of Flexible Dynamic Quantile Linear Models}

\Plainauthor{Antonio De Leon, Raquel Barata, Raquel Prado, Bruno Sans\'{o}}
\Plaintitle{exdqlm: An R Package for Estimation and Analysis of Flexible Dynamic Quantile Linear Models}
\Shorttitle{\pkg{exdqlm}: Flexible Dynamic Quantile Linear Models}

\Abstract{ We present the \proglang{R} package \pkg{exdqlm} for Bayesian quantile regression, with primary emphasis on dynamic state-space quantile models for time series. The package is built around extended dynamic quantile linear models (exDQLMs), which use the extended asymmetric Laplace (exAL) family, a parametric extension of the asymmetric Laplace (AL) distribution commonly used in quantile regression. The software provides posterior simulation via Markov chain Monte Carlo (MCMC) and fast approximate posterior inference via Laplace--delta variational Bayes (LDVB), supporting posterior uncertainty quantification while also providing a computationally efficient option for longer time series. The same package interface supports static exAL quantile regression with regularized priors, dynamic transfer-function models for nonlinear input effects at a given quantile, post hoc posterior-predictive synthesis across separately fitted quantiles, forecasting, and quantitative and visual diagnostics for model evaluation. }


\Address{
  Antonio De Leon and Raquel Barata\\
  Department of Statistics\\
  University of California Santa Cruz\\
  Santa Cruz, CA 95064\\
  E-mail: \email{jaguir26@ucsc.edu}, \email{raquel.a.barata@gmail.com}\\
}

\usepackage{Sweave}
\RecustomVerbatimEnvironment{Code}{Verbatim}{fontsize=\footnotesize}
\RecustomVerbatimEnvironment{CodeInput}{Verbatim}{fontsize=\footnotesize,fontshape=sl}
\RecustomVerbatimEnvironment{CodeOutput}{Verbatim}{fontsize=\footnotesize}
\RecustomVerbatimEnvironment{Sinput}{Verbatim}{fontsize=\footnotesize,fontshape=sl}
\RecustomVerbatimEnvironment{Soutput}{Verbatim}{fontsize=\footnotesize}
\RecustomVerbatimEnvironment{Scode}{Verbatim}{fontsize=\footnotesize,fontshape=sl}

\begin{document}

\section{Introduction}

Quantile regression has become a widely used alternative to mean-based modeling in static settings. However, dynamic quantile methods and accompanying software remain comparatively limited. In Bayesian parametric quantile regression, the asymmetric Laplace (AL) likelihood has played a central role because it is linked to the check-loss function used in classical quantile regression \citep{yu2001bayesian,koenkerquantile} and allows convenient mixture representations for posterior computation. Nevertheless, the AL distribution can be restrictive as an error model. More flexible error distributions have been studied extensively in the Bayesian nonparametric literature \citep{kottas2009bayesian,reich2009flexible,taddy2010bayesian}; however, the resulting methods can be computationally demanding for dynamic models and longer time series.

\citet{yan2025family} introduce the extended asymmetric Laplace (exAL) distribution, a parametric extension of the AL distribution that preserves many of its computational advantages while relaxing some restrictive assumptions. Their work also develops Bayesian exAL regression for static quantile models. Building on the exAL family, \citet{me} developed the extended dynamic quantile linear model (exDQLM) framework for modeling time-varying quantiles, together with Markov chain Monte Carlo (MCMC) and importance-sampling variational Bayes (ISVB) inference and a transfer-function extension for lagged covariate effects. The Laplace--delta variational Bayes (LDVB) implementation considered here follows the general nonconjugate variational inference strategy of \citet{wang2013nonconjugatevb}, specialized to the exAL scale--skewness block. Thus, the exAL distribution, the exDQLM state-space model formulation, the MCMC and ISVB algorithms in \citet{me}, and the general Laplace--delta variational strategy constitute the principal methodological foundations behind the present work.

The contribution of the present article is software-centered. The \proglang{R} package \pkg{exdqlm} turns these methodological components into an integrated workflow for specifying, fitting, inspecting, diagnosing, forecasting, and reproducing Bayesian dynamic and static quantile analyses. The package makes five main software contributions:
\begin{itemize}
\item a composable model-construction interface for trend, seasonal, regression, and transfer-function components in Bayesian dynamic quantile state-space models;
\item coordinated MCMC and variational inference interfaces, including LDVB and ISVB, for dynamic and static fits;
\item an S3 object workflow for fitted models, post-processing and forecasting objects with \code{print()}, \code{summary()}, \code{plot()}, \code{predict()}, and \code{diagnostics()} methods where appropriate;
\item package-level exAL distribution utilities, predictive scoring and calibration diagnostics, and compiled implementations of computationally intensive model-building, posterior sampling and posterior predictive operations, with \proglang{R} fallbacks where applicable; and
\item support for static AL/exAL regression with shrinkage-prior options, together with scripts and documentation that reproduce the analyses reported in this article.
\end{itemize}

The \proglang{R} \citep{R} software ecosystem for quantile regression is broad, but the capabilities most relevant to this paper are distributed across software with different aims. Table \ref{table:software} compares representative packages along four traits that are key to the scope of \pkg{exdqlm}: the primary modeling objective, the treatment of the temporal or latent-state structure, the inference strategy, and handling of multiple quantile levels.
The distinction between time included as an observed predictor and coefficients propagated as latent states is particularly important.
The computation column reports how fitting, posterior approximation, or state summaries are obtained, rather than treating these as additional model structures.
\begin{table}[h!]
\centering
\begingroup
\CompactTableStyle
\begin{tabularx}{\textwidth}{@{}l@{\hspace{0.7em}} >{\raggedright\arraybackslash}X >{\raggedright\arraybackslash}X >{\raggedright\arraybackslash}X >{\raggedright\arraybackslash}X@{}}
\toprule
\textbf{Package} & \textbf{Main scope} & \shortstack{\textbf{Temporal or}\\\textbf{state-space handling}} & \shortstack{\textbf{Inference or}\\\textbf{computation}} & \shortstack{\textbf{Multiple-quantile}\\\textbf{handling}} \\
\midrule
\pkg{quantreg} & General frequentist QR & Time-series regression formula: \code{dynrq()} handles lags, differences, trends, and seasons; no latent states & Optimization-based fitting; frequentist and bootstrap summaries & Vector-valued \code{tau}; post hoc rearrangement \\
\addlinespace[1pt]
\pkg{bayesQR} & Bayesian AL QR & Time or lags only as user-supplied covariates; no serial or state-space model & MCMC under the AL likelihood; optional adaptive lasso & Vector-valued quantiles; output by fitted quantile \\
\addlinespace[1pt]
\pkg{pqrBayes} & Bayesian penalized QR & Varying-coefficient regression: coefficients are B-spline functions of an observed modifier, possibly time; no serial or state-space model & MCMC with shrinkage priors & One target quantile per fit \\
\addlinespace[1pt]
\pkg{brms} & General Bayesian regression with an AL family & Regression autocorrelation terms can be added; no quantile state-space workflow & Stan/NUTS posterior sampling & Fixed through the \code{asym\_laplace()} quantile parameter \\
\addlinespace[1pt]
\pkg{qgam} & Smooth additive QR & Time may enter as a smooth covariate; no serial or state-space model & Calibrated Gibbs posterior with GAM fitting & Single quantile via \code{qgam()}; multiple via \code{mqgam()} \\
\addlinespace[1pt]
\pkg{qrjoint} & Bayesian joint linear QR & Time only as a predictor; no serial or state-space model & Adaptive MCMC for joint quantile planes & Joint noncrossing quantile planes \\
\addlinespace[1pt]
\pkg{dlm} & Gaussian state-space modeling & Gaussian latent states with filtering, smoothing, and forecasting; not QR & Kalman state recursions; MLE or Bayesian parameter analysis & Not a quantile-regression package \\
\addlinespace[1pt]
\midrule
\pkg{exdqlm} & Bayesian exAL QR & Quantile latent states with filtering/smoothing summaries, forecasting, and transfer functions & MCMC posterior simulation; VB approximate posterior inference & Post hoc synthesis from separately fitted quantiles \\
\bottomrule
\end{tabularx}
\endgroup
\caption{\proglang{R} packages relevant to quantile and state-space modeling. Entries summarize documented primary interfaces along the dimensions used to position \pkg{exdqlm}; they are selective rather than exhaustive. In the temporal column, time-as-predictor/smooth/modifier uses, regression autocorrelation, Gaussian latent states, and quantile latent states are distinct categories. QR denotes quantile regression, AL asymmetric Laplace, exAL its extended version, NUTS the No-U-Turn sampler, GAM generalized additive model, MLE maximum likelihood estimation, and VB variational Bayes.}
\label{table:software}
\end{table}

The frequentist package \pkg{quantreg} \citep{quantreg} remains the broad general-purpose baseline, covering linear, nonlinear, nonparametric, censored, and dynamic quantile regression, together with bootstrap-based and related uncertainty summaries. Its \code{dynrq()} function supports dynamic linear quantile regression through time-series regression formulas involving lags, differences, trends, seasonal factors, and harmonic terms. However, the dynamic structure enters through the design matrix rather than through latent states.
Once these regressors have been constructed, estimation proceeds through the standard \pkg{quantreg} fitting routines. Thus, \code{dynrq()} is a regression-style time-series interface rather than one for a latent state-space model.
Most other \proglang{R} packages relevant here implement static quantile regression or target a different multi-quantile objective. Among the static Bayesian options, \pkg{bayesQR} \citep{bayesQR} provides Bayesian quantile regression under the AL working likelihood with an adaptive-lasso option, whereas \pkg{pqrBayes} \citep{pqrBayes} emphasizes penalized Bayesian quantile regression with spike-and-slab and horseshoe-family shrinkage priors. The varying-coefficient option, specified by \code{model = "VC"} in \pkg{pqrBayes}, models each regression coefficient as a smooth function of a user-supplied observed modifier. It represents these functions using a B-spline basis and estimates the corresponding basis coefficients using MCMC under shrinkage priors. Thus, \pkg{pqrBayes} is appropriate for static varying-coefficient quantile regression, including applications where the chosen modifier is calendar time or another time index. Even in such applications, however, the coefficients are modeled as smooth functions of time rather than propagated through a latent state equation.  The broader Bayesian modeling package \pkg{brms} \citep{brmsfamily} is also relevant because it supports AL-based quantile regression within a flexible multilevel, nonlinear, and additive regression framework, and can incorporate regression-level autocorrelation terms. However, it is not organized as a dedicated quantile state-space workflow, and its documentation does not present filtered or smoothed state summaries of the type targeted by \pkg{exdqlm}. Packages such as \pkg{qgam} \citep{qgam}, \pkg{qrjoint} \citep{qrjoint}, \pkg{quantregGrowth} \citep{quantregGrowth}, and \pkg{qrcm} \citep{qrcm} address additive, joint, non-crossing, or coefficient-modeling formulations. Other packages focus on specialized static settings, including penalized frequentist fitting across quantiles in \pkg{rqPen} \citep{rqPen}, convolution-smoothed linear quantile regression with efficient gradient-based fitting in \pkg{conquer} \citep{conquer}, and robust or censored-response modeling in \pkg{lqr} \citep{lqr}.

Quantile-regression software implementations are also available in Python, MATLAB and Julia, but they are general static quantile regression tools and do not provide the Bayesian latent-state quantile workflow considered here.
\pkg{exdqlm} is not intended to replace broad general-purpose quantile-regression toolkits, additive quantile-regression packages, or software designed specifically for joint multi-quantile estimation. Its role is more specific: it fills a gap between general quantile-regression software and general Gaussian state-space software by providing a dedicated Bayesian workflow for dynamic quantile state-space models. General state-space packages in \proglang{R}, such as \pkg{dlm} \citep{dlm} and \pkg{dynr} \citep{dynr}, remain useful for dynamic modeling, but they do not provide the quantile likelihoods, filtered and smoothed quantile-state summaries, diagnostics, forecasting, or synthesis tools targeted here.

The \code{exdqlm} package is distributed under the MIT License and is available from the Comprehensive R Archive Network at \url{https://CRAN.R-project.org/package=exdqlm}. The package source corresponding to this article is version 1.1.0.

The remainder of the paper is organized as follows. Section \ref{sec:exdqlm} summarizes the modeling framework, including algorithms used for posterior inference and transfer-function state augmentation for nonlinear input effects. It also discusses static quantile regression as a special case of the framework. Section \ref{sec:diags} details the model diagnostics and forecasting method implemented in the package. Section \ref{sec:design} describes the package design and implementation, including the object families and standard methods used by the examples. Section \ref{sec:exs} illustrates the package through four examples. Section \ref{sec:summary} summarizes the package capabilities, and Appendices A--F collect selected posterior targets, package-specific computational blocks, diagnostic formulas, and predictive-synthesis details used by the implementation.


\section{Extended dynamic quantile linear models}
\label{sec:exdqlm}
This section provides an overview of extended dynamic quantile linear models (exDQLMs) along with related posterior inference and tools for model diagnostics. Additional implementation details and posterior derivations are provided in Appendices A--F.
\subsection{General modeling framework}
\subsubsection{Model formulation}
Let \(y_t\) denote a scalar response observed at times \(t=1,\ldots,T\). For each $t$, an extended dynamic quantile linear model (exDQLM) for the $p_0$ quantile is defined by
\begin{align}
\label{EQ:exDLM}
& \text{Observation equation:} & y_t &= \mat{F}_t^\top \vect{\theta}_t + \epsilon_t,  & \epsilon_t & \sim \exAL_{p_0}(0,\sigma,\gamma) \\
\label{EQ:exDLMevo}
& \text{System equation:} & \vect{\theta}_t &= \mat{G}_t \vect{\theta}_{t-1} + \vect{\omega}_t, & \vect{\omega}_t  & \sim \N(\mat{0},\mat{W}_t).
\end{align}
Here $\mat{F}_t$ is the $q \times 1$ regression vector of the covariates, $\vect{\theta}_t$ is the corresponding $q \times 1$ latent state vector and $\mat{G}_t$ is the $q\times q$-dimensional evolution matrix. The state innovation $\vect{\omega}_t$ follows a $q$-variate normal distribution with state covariance matrix $\mat{W}_t.$
The observational errors $\epsilon_t$ follow an extended asymmetric Laplace (exAL) distribution for fixed quantile $p_0$, whose density is denoted as $\exAL_{p_0}(\cdot)$, and it is such that $\int_{-\infty}^0 \exAL_{p_0}(\epsilon_t |0,\sigma,\gamma) \mathrm{d}\epsilon_t = p_0$. Thus, the observation equation in \eqref{EQ:exDLM} implies that $\mat{F}_t^\top \vect{\theta}_t$ corresponds to the $p_0$-quantile of $y_t$. Table \ref{table:priors} shows how the state-space quantities and prior hyperparameters  are represented in the \pkg{exdqlm} function inputs.

The exAL distribution, introduced by \citet{yan2025family}, extends the location-scale mixture representation of the AL \citep{kozumi2011gibbs}. For fixed $p_0$, the skewness parameter $\gamma$ is restricted to an interval $(L,U)$, where $L$ and $U$ are the negative and positive roots of $g(\gamma) = 1-p_0$ and $g(\gamma) = p_0$, respectively, with $g(\gamma) = 2\Phi(-|\gamma|)\exp(\gamma^2/2)$. This yields the following mixture representation of $\exAL_{p_0}(0,\sigma,\gamma)$:
\begin{equation}
\label{EQ:exALmix}
\begin{aligned}
\exAL_{p_0}(y_t \mid 0,\sigma,\gamma)
&=
\int\int_{\mathbb{R}^+ \times \mathbb{R}^+}
\N(y_t \mid C(p,\gamma) \sigma |\gamma| s_t + A(p) v_t, \sigma B(p) v_t)\\
&\quad \times \Exp(v_t \mid \sigma) \N^+(s_t \mid 0,1) \mathrm{d}v_t \mathrm{d}s_t.
\end{aligned}
\end{equation}
Here, $p = p(p_0,\gamma) = I(\gamma <0) + \{[p_0 - I(\gamma <0)]/g(\gamma) \}$ where $g(\gamma) = 2\Phi(-|\gamma|)\exp(\gamma^2/2)$, $\Phi(\cdot)$ denotes the standard normal CDF, and $I(\cdot)$ is the indicator function. $A(p)$, $B(p)$, $C(p,\gamma)$ are the functions of $p$ and $\gamma$: $A(p) = \frac{1-2p}{p(1-p)}$, $B(p) = \frac{2}{p(1-p)}$, $C(p,\gamma) = [I(\gamma>0) -p]^{-1}$. Lastly, $\Exp(v \mid \sigma)$ denotes the exponential distribution with mean $\sigma$, and $\N^+(s_t \mid 0,1)$ denotes a normal distribution truncated to the positive reals with mean $0$ and variance $1$. Introducing the latent variables $v_t$ and $s_t$ gives the hierarchical representation:
\begin{align}
\label{EQ:exDQLMobs}
y_t | \vect{\theta}_t,\sigma,\gamma,v_t,s_t  & \sim \N(y_t \mid \mat{F}_t^\top \vect{\theta}_t + C(p,\gamma) \sigma |\gamma| s_t + A(p) v_t,\sigma B(p) v_t) \\
v_t, s_t | \sigma & \sim \Exp(v_t \mid \sigma) \N^+(s_t \mid 0,1) \\
\label{EQ:exDQLMevo}
\vect{\theta}_t | \vect{\theta}_{t-1},\mat{W}_t & \sim  \N(\mat{G}_t \vect{\theta}_{t-1},\mat{W}_t).
\end{align}
The posterior algorithms in \pkg{exdqlm} use this hierarchical form.


\subsubsection{Prior specification}
\label{sec:priors}

Conditional on $p_0$, the model assigns priors to the initial state $\vect{\theta}_0$, scale parameter $\sigma$ and skewness parameter $\gamma$. For $\vect{\theta}_0$, we assume a $q$-variate conjugate normal prior $\vect{\theta}_0 \sim \N(\mat{m}_0,\mat{C}_0)$ and the scale parameter follows a conjugate inverse-gamma prior, $\sigma \sim \IG(a_{\sigma},b_{\sigma})$.
The parameter $\gamma$ has bounded support over the interval $(L,U)$ where $L$ and $U$ are the negative and positive roots associated with $p_0$ in the exAL representation\citep{yan2025family}. The package uses a proper Student-\(t\) prior on the skewness parameter, truncated to the admissible interval \((L,U)\), i.e., $\gamma \sim \text{t}_{(L,U)}(m_\gamma,s_\gamma)$ with $\nu_{\gamma}$ degrees of freedom, which improves numerical stability while respecting the support restrictions from the exAL family. Table \ref{table:priors} gives the corresponding package inputs and default values.

\begin{table}[!htb]
\centering
\TableStyle
\begin{tabular}{@{}lccccccccc@{}}
\toprule
\proglang{R} input & \multicolumn{4}{c}{\code{model}} & \multicolumn{2}{c}{\code{PriorSigma}}    & \multicolumn{3}{c}{\code{PriorGamma}}  \\
\cmidrule(lr){2-5}\cmidrule(lr){6-7}\cmidrule(l){8-10}
Mathematical symbol & $\mat{F}_{1:T}$ & $\mat{G}_{1:T}$ & $\mat{m}_0$ & $\mat{C}_0$ & $a_{\sigma}$ & $b_{\sigma}$ & $m_{\gamma}$ & $s_{\gamma}$ & $\nu_{\gamma}$ \\
\proglang{R} component & \code{FF} & \code{GG} & \code{m0} & \code{C0} & \code{a_sig} & \code{b_sig} & \code{m_gam} & \code{s_gam} & \code{df_gam} \\
Default if omitted & -- & -- & $\mat{0}_q$ & $10^3\mat{I}_q$ & 2.1 & 1.1 & 0 & 1 & 1 \\
\bottomrule
\end{tabular}
\caption{Translation from exDQLM notation to \pkg{exdqlm} inputs. The \code{model} object stores state-space components and the initial-state prior, while \code{PriorSigma} and \code{PriorGamma} store hyperparameters for $\sigma$ and $\gamma$. A dash indicates no generic default because \code{FF} and \code{GG} are determined by the selected model components. Here $\mat{0}_q$ and $\mat{I}_q$ denote a $q$-vector of zeros and the $q$-dimensional identity matrix.}
\label{table:priors}
\end{table}

\subsubsection{Posterior estimation}

The dynamic exDQLM posterior target and the original MCMC and ISVB inference strategy were developed by \citet{me}. The package makes \code{exdqlmISVB()} available for variational inference following the ISVB approach of \citet{me}.
The principal fitting routines in the package are
\code{exdqlmMCMC()}, which performs posterior inference via MCMC, and \code{exdqlmLDVB()}, which provides a faster variational approximation based on
the Laplace--delta strategy of \citet{wang2013nonconjugatevb}.

The MCMC routine uses Gibbs sampling steps to update the latent-variable sequences
$\{s_t\}_{t=1}^T$ and $\{v_t\}_{t=1}^T$. The dynamic state sequence $\vect{\theta}_{1:T}$ is sampled using a forward-filtering backward-sampling algorithm \citep{carter1994gibbs,fruhwirth1994data}.
The scale and skewness parameters are updated separately from the latent-state blocks. By default, the package samples $\sigma$ from its conditional distribution given $\gamma$ and then updates $\gamma$ using a bounded univariate slice sampler \citep{neal2003slice}. Joint
random-walk Metropolis--Hastings (MH) alternatives are also available.
The chain can optionally be initialized using the variational
approximation, but this initialization is separate from the subsequent MCMC
update kernel.

For longer time series or repeated model comparisons, MCMC can be computationally demanding. The \code{exdqlmLDVB()} routine uses a mean-field factorization and blockwise coordinate-ascent variational inference. The blocks associated with the latent variables and dynamic states admit closed-form coordinate updates, whereas the joint scale–skewness block is non-conjugate. Following \citet{wang2013nonconjugatevb}, this block is approximated on transformed coordinates, and a second-order delta method evaluates the expectations needed by the remaining updates. The dynamic states are updated at each iteration using forward-filtering backward-smoothing, and convergence is monitored through changes in the state, scale, skewness, and evidence lower bound (ELBO).

Relative to the ISVB implementation, LDVB replaces the stochastic importance-sampling approximation for the non-conjugate scale–skewness block with a deterministic optimization-based approximation conditional on the data and tuning inputs. This leads to reproducible convergence monitoring and provides approximate uncertainty summaries for both $\sigma$ and $\gamma$. After convergence, samples from the fitted variational factors are generated for posterior summarization. These are approximate variational draws rather than draws from the exact posterior, so MCMC remains the reference calculation when posterior tail behavior or interval accuracy is central.

Table \ref{table:post} shows selected core inference controls for the fitting functions \code{exdqlmMCMC()} and \code{exdqlmLDVB()}.
The MCMC routine also accepts \code{Sig.mh} as an optional proposal covariance when \code{mh.proposal} selects a joint random-walk MH kernel; it is not used under the default slice-sampling update.
Both fitting routines also use package-level options to control selected compiled and R backend paths. These runtime options, including ELBO monitoring where applicable, are described in Section
 \ref{sec:design} and are distinct from the function-specific inference controls summarized in Table \ref{table:post}.

\begin{table}[!htb]
\centering
\TableStyle
\begin{tabular}{@{}lccccc@{}}
 \toprule
 Control & \multicolumn{3}{c}{\code{exdqlmMCMC()}} & \multicolumn{2}{c}{\code{exdqlmLDVB()}}  \\
 \cmidrule(lr){2-4}\cmidrule(l){5-6}
 Mathematical symbol & $\mathcal{K}_{\sigma,\gamma}$ & $N_{BURN}$ & $N_{MCMC}$ & $\epsilon_{tol}$ & $N_{SAMP}$\\
 \midrule
 \proglang{R} argument & \code{mh.proposal} & \code{n.burn} & \code{n.mcmc} & \code{tol} & \code{n.samp} \\
 Default if omitted & \code{"slice"} & 2000 & 1500 & 0.1 & 200 \\
 \bottomrule
\end{tabular}
\caption{Core inference controls for \code{exdqlmMCMC()} and \code{exdqlmLDVB()}. $\mathcal{K}_{\sigma,\gamma}$ denotes the update kernel for the $(\sigma,\gamma)$ block, $N_{BURN}$ is the burn-in length, $N_{MCMC}$ is the number of retained MCMC draws, $\epsilon_{tol}$ is the LDVB convergence tolerance and $N_{SAMP}$ is the number of requested variational draws.
Advanced control-list arguments and backend options are documented separately.}
\label{table:post}
\end{table}


\subsubsection{Evolution covariance and discount factors}
\label{sec:dfs}

Both \code{exdqlmMCMC()} and \code{exdqlmLDVB()} use discount factors to specify the evolution covariance matrices $\mat{W}_t$ in Equation \eqref{EQ:exDLMevo} \citep[see][and references therein]{west1997bayesian,prado_ferreira_west2021}. Under a common discount factor $\delta \in (0,1],$ $\mat{W}_t = \frac{1-\delta}{\delta} \mat{G}_t \mat{C}_{t-1} \mat{G}_t^\top$ where $\mat{C}_{t-1}$ denotes the posterior variance of the state vector $\vect{\theta}_{t-1}$ at time $t-1$. The discount factor $\delta$ can be specified by the user via the function parameter \code{df}. The selection of discount factors can be done by optimizing a given selection criterion such as the continuous ranked probability score (CRPS) or the posterior predictive loss criterion (PPLC) defined in Section \ref{sec:diags}, with Kullback--Leibler (KL) divergence as a complementary calibration diagnostic.  Discount factor selection is illustrated in Section \ref{sec:ex2}.

Component-specific discounting is also available. When the state vector is partitioned into $h$ components with dimensions $q_1,\ldots,q_h$, each block of the evolution covariance matrix can be assigned its own discount factor.
The arguments \code{df} and \code{dim.df} specify the component discount factors $(\delta_1,\dots,\delta_h)$ and their corresponding dimensions $(q_1,\dots,q_h)$.  Section \ref{sec:exs} provides examples.  \cite{west1997bayesian} and \cite{prado_ferreira_west2021} give a detailed treatment of component discounting.


\subsubsection{Transfer function models}
\label{sec:tf}

Transfer functions provide a state-space representation for current and lagged input effects on a response quantile.
In \pkg{exdqlm}, this is handled by augmenting the dynamic state-space representation conditional on a fixed transfer-function rate \(\lambda\), and the resulting workflow is available through both LDVB and MCMC fitting routines.
For a transfer input vector $\mat{x}_t \in \mathbb{R}^k$, a transfer-function exDQLM with exponential decay is defined by
\begin{align}
\label{EQ:begin_transfer}
y_t | \vect{\theta}_t,\zeta_t,\gamma,\sigma  & \sim \exAL_{p_0}(\mat{F}_t^\top \vect{\theta}_t +  \zeta_{t},\sigma,\gamma) \\
\vect{\theta}_t | \vect{\theta}_{t-1},\mat{W}_t^\theta  & \sim  \N(\mat{G}_t \vect{\theta}_{t-1},\mat{W}_t^\theta) \\
\label{EQ:transfer_effect}
\zeta_{t}|\zeta_{t-1},\vect{\psi}_{t-1},\omega_{t} & \sim \N(\lambda \zeta_{t-1} + \mat{x}_{t}^\top \vect{\psi}_{t-1},\omega_{t}) \\
\label{EQ:end_transfer}
\vect{\psi}_{t}|\vect{\psi}_{t-1},\mat{W}_{t}^{\psi} & \sim \N(\vect{\psi}_{t-1},\mat{W}_{t}^{\psi}).
\end{align}
Here $\zeta_t \in \mathbb{R}$ denotes the accumulated transfer effect on the quantile at time $t$ and $\vect{\psi}_t \in \mathbb{R}^k$ is the vector of instantaneous transfer coefficients associated with the components of $\mat{x}_t$. The mean contribution of the transfer block is therefore $\lambda \zeta_{t-1} + \mat{x}_t^\top \vect{\psi}_{t-1}$: the inner product $\mat{x}_t^\top \vect{\psi}_{t-1}$ captures the contemporaneous effect of the current inputs, and $\lambda\in(0,1)$ controls the geometric decay of previous accumulated effects. More explicitly, the contribution of $\mat{x}_t$ to the quantile at time $t+h$ is $\lambda^h a_t$, where \(a_t=\mat{x}_t^\top \vect{\psi}_{t-1}\). For tolerance $\epsilon>0$, the package summarizes the continuous decay horizon
\[
k_t =
\begin{cases}
\{\log(\epsilon)-\log(|a_t|)\}/\log(\lambda), & |a_t|>\epsilon,\\
0, & |a_t|\leq \epsilon,
\end{cases}
\]
whose positive values solve \(\lambda^{k_t}|a_t|=\epsilon\). The reported \code{median.kt} is the median of these horizons over time for $\epsilon = 1\mathrm{e}{-3}$.

Conditional on a fixed value $\tilde\lambda$, the dynamic fitting routines incorporate the transfer-function structure by replacing $\mat{F}_t$, $\vect{\theta}_t$, $\mat{G}_t$, and $\mat{W}_t^\theta$ in Equations \eqref{EQ:exDLM}-\eqref{EQ:exDLMevo} with
\[
\tilde{\mat{F}}_t^\top = (\mat{F}_t^\top, 1, \mat{0}_k^\top), \qquad
\tilde{\vect{\theta}}_t^\top = (\vect{\theta}_t^\top,\zeta_t,\vect{\psi}_t^\top),
\]
\[
\tilde{\mat{G}}_t = \operatorname{blockdiag}\{\mat{G}_t, \mat{G}^{tf}_t\},
\qquad
\mat{G}^{tf}_t =
\begin{pmatrix}
\tilde\lambda & \mat{x}_t^\top \\
\mat{0}_k & \mat{I}_k
\end{pmatrix},
\]
and $\tilde{\mat{W}}_t = \operatorname{blockdiag}\{\mat{W}_t^\theta,\omega_t,\mat{W}_t^\psi\}.$
This augmentation is implemented in both functions \code{exdqlmTransferLDVB()} and \code{exdqlmTransferMCMC()}.
Discount factors are used to specify the transfer block variances $\omega_t$ and $\mat{W}_t^\psi$.
In the package, \code{tf.df} can be supplied as a single shared discount factor, as a pair $(\delta_\zeta,\vect{\delta_\psi})$ with a shared value for the whole $\vect{\psi}_t$ block, or componentwise across $(\zeta_t,\psi_{1,t},\dots,\psi_{k,t})$. A conjugate normal prior on $(\zeta_{0},\vect{\psi}_{0}^\top)^\top \sim \N(\mat{m}_0^{tf},\mat{C}_0^{tf})$, with $\mat{m}_0^{tf}\in\mathbb{R}^{k+1}$ and $\mat{C}_0^{tf}\in\mathbb{R}^{(k+1)\times(k+1)}$, completes the transfer-function model extension. Table \ref{table:tf} shows the transfer-function arguments.

\begin{table}[!htb]
\centering
\TableStyle
\begin{tabularx}{\textwidth}{@{}>{\raggedright\arraybackslash}p{0.15\textwidth}>{\centering\arraybackslash}p{0.15\textwidth}>{\raggedright\arraybackslash}p{0.14\textwidth}>{\raggedright\arraybackslash}X@{}}
	\toprule
	Quantity & Mathematical symbol & \proglang{R} argument & Input/default \\
	\midrule
	Transfer inputs & $\mat{x}_t$ & \code{X} & Required numeric vector or $T\times k$ matrix. \\
	Transfer rate & $\tilde\lambda$ & \code{lam} & Required scalar in $(0,1)$. \\
	Discount factors & $(\delta_\zeta,\vect{\delta_\psi})$ & \code{tf.df} & Required vector: length 1 (shared); length 2 ($\delta_\zeta$ and shared $\vect{\delta_\psi}$); or length $k+1$ (componentwise). \\
	Prior mean & $\mat{m}_0^{tf}$ & \code{tf.m0} & Optional vector of length $k+1$; default $\mat{0}_{k+1}$. \\
	Prior covariance & $\mat{C}_0^{tf}$ & \code{tf.C0} & Optional $(k+1)\times(k+1)$ covariance matrix; default $\mat{I}_{k+1}$. \\
	\bottomrule
\end{tabularx}
\caption{Transfer-function inputs for \code{exdqlmTransferLDVB()} and \code{exdqlmTransferMCMC()}. Here $k$ denotes the number of transfer covariates.}
\label{table:tf}
\end{table}


\subsubsection{Special cases}
\label{sec:special}

Several commonly used models are special cases of the \code{exDQLM}. First, setting $\gamma=0$ reduces the exAL observation distribution to the AL distribution, i.e.,
$\epsilon_t \sim \exAL_{p_0}(0,\sigma,\gamma=0) = \AL_{p_0}(0,\sigma),$ leading to the dynamic quantile linear model (DQLM) of \cite{gonccalves2017dynamic}. The DQLM can be implemented with our package using the parameter \code{dqlm.ind = TRUE}.

Second, setting $\mat{G}_t=\mat{I}$ and $\mat{W}_t=\mat{0}$ reduces the model to the static exAL regression of \cite{yan2025family}.
This model can be implemented using functions \code{exalStaticMCMC()} or \code{exalStaticLDVB()}.

Finally, combining these two restrictions gives static Bayesian linear quantile regression under the AL working likelihood \cite{yu2001bayesian}, as implemented in packages such as \pkg{bayesQR}. Within the \pkg{exdqlm} package, this model can be fitted using \code{exalStaticMCMC()} or \code{exalStaticLDVB()} with \code{al.ind = TRUE}, a static convenience alias for \code{dqlm.ind = TRUE}, which fixes $\gamma=0$.
Examples of these special models are given in Section \ref{sec:exs}, with the sparse static regression workflow illustrated in Section \ref{sec:ex4static}.


\subsection{Model diagnostics and forecasting}
\label{sec:diags}

The package includes quantitative and graphical diagnostics for checking fitted quantile calibration, predictive performance, and forecasts.
Calibration is assessed using plug-in one-step-ahead probability integral transforms and standardized forecast errors. Predictive performance is summarized using CRPS and a target-quantile posterior-predictive loss criterion, while future quantiles and optional posterior-predictive draws are obtained through the standard \code{predict()} interface.

\subsubsection{Calibration diagnostics}
The one-step-ahead predictive probability integral transform (PIT) introduced by \cite{rosenblatt1952remarks} is a useful diagnostic for models in which the marginal forecast distributions are not available in closed-form \citep{huerta2003time,prado2006multivariate}. Let $Y_t$ be the random response at time $t$ with observed value $y_t$ and let $y_{1:(t-1)}$ denote the observations available up to time $t-1.$ The one-step-ahead PIT is
\begin{align}
u_t^\star = \Pr(Y_t \leq y_t \mid y_{1:t-1}),
\label{eq:pit_ideal}
\end{align}
 Under the true conditional distribution of $Y_t$
  given the past at every time, the PIT values $\{u_t^*\}$ are independent Uniform$(0,1)$ random variables \citep{rosenblatt1952remarks}.

  The package reports a maximum a posteriori (MAP) plug-in version based on the conditional Gaussian filtering representation. If \(\hat f_t\) and \(\hat q_t\) denote the MAP plug-in one-step-ahead forecast location and variance, the stored standardized forecast error and the corresponding PIT diagnostic are
\begin{align}
\hat e_t =
\frac{y_t-\hat f_t}{\sqrt{\hat q_t}}, \;\;\; \text{and}
\label{eq:standardized_error}
\;\;\;
\hat{u}_t = \Phi(\hat e_t),
\end{align}
where $\Phi$ denotes the standard normal CDF. These are diagnostic summaries of the fitted one-step-ahead sequence rather than posterior uncertainty bands for residuals.
Serial dependence in $\{\hat u_t\}$ is examined using an autocorrelation function (ACF) plot. The normal transformation $\Phi^{-1}(\hat{u}_t),$ which corresponds to $\hat{e}_t$ under the plug-in Gaussian representation is assessed using a normal quantile--quantile (QQ) plot.

Departure from normality is also summarized using the KL divergence \citep{kullback1951information}.
If $P_e$ denotes the density estimate of the MAP standardized forecast-error distribution and $\phi$ denotes the standard normal density, the package reports a forward diagnostic, denoted $\KL(P_e\,|| \ \phi)$ and, as a sensitivity measure, the reversed diagnostic $\KL(\phi \,||\,P_e)$.
Smaller values suggest better calibration.
These MAP standardized forecast errors are included in the output of the \code{exdqlmMCMC()} and \code{exdqlmLDVB()} functions.

\subsubsection{Predictive scoring}
The continuous ranked probability score (CRPS) evaluates the full posterior predictive distribution \citep{gneiting2007}.
Let $\rho_p(x)=x[p-I(x<0)]$ denote the check loss at level $p$. For a predictive distribution $G$ and observation $y$, the CRPS is
\begin{equation}
\CRPS(y,G)=\mathbb{E}_{Y\sim G}|Y-y|-\frac{1}{2}\mathbb{E}_{Y,Y'\sim G}|Y-Y'|
=2\int_0^1 \rho_p\!\left(y-G^{-1}(p)\right)\,dp,
\label{eq:crps_main}
\end{equation}
which shows that CRPS integrates check loss across all quantile levels and therefore evaluates the full predictive distribution rather than evaluating a single target level \(p_0\). In \code{diagnostics()}, posterior predictive draws define empirical quantiles \(\widehat q_t(\tau_k)\), giving the finite approximation
\begin{equation}
\widehat{\CRPS}_t
=
2\sum_{k=1}^{K} w_k
\rho_{\tau_k}\{y_t^{obs}-\widehat q_t(\tau_k)\},
\label{eq:crps_iqs_main}
\end{equation}
where the weights \(w_k\) sum to one \citep{laio2007verification}. The package uses the grid \(\tau_k=0.01,0.02,\ldots,0.99\) with equal weights and reports the CRPS values in \eqref{eq:crps_iqs_main} averaged over time. Smaller CRPS values indicate better calibrated models and sharper predictive performance. For a deterministic forecast, CRPS reduces to the absolute error, so its sample mean CRPS equals the mean absolute error (MAE).

Motivated by the posterior-predictive loss (PPL) framework of \cite{gelfand1998model}, the package returns \code{pplc}, which evaluates the posterior predictive distribution under the target-level check loss $\rho_{p_0}$. That is,
\begin{align}
\PPLC = \sum_t \E[\rho_{p_0} (y_t^{obs}-y^{rep}_t) \mid y_{1:T}],
\label{eq:pplc_main}
\end{align}
where the expectation is with respect to posterior predictive replicate draws generated from the observation equation \eqref{EQ:exDLM} under posterior draws of the model parameters.
Smaller PPLC values indicate better performance.

Samples from the posterior replicate distributions are included in the output of \code{exdqlmMCMC()} and \code{exdqlmLDVB()}.

Then, the function \code{diagnostics()} computes these calibration and predictive summaries for fitted dynamic objects. It returns an \code{exdqlmDiagnostic} object that can be inspected with \code{print()} or \code{summary()} and visualized with \code{plot()}.

\subsubsection{k-step-ahead forecasting}

For an arbitrary time $\tilde{t}$, the $k$-step-ahead future marginal distribution of the quantile is

\begin{equation}
\mat{F}_{\tilde{t}+k}^\top \vect{\theta}_{\tilde{t}+k}\mid y_1,\dots,y_{\tilde{t}} \sim \N(\mat{F}_{\tilde{t}+k}^\top \mat{a}_{\tilde{t}}(k),\mat{F}_{\tilde{t}+k}^\top\mat{R}_{\tilde{t}}(k)\mat{F}_{\tilde{t}+k})
\label{eq:forecast_recursion}
\end{equation}

where $\mat{a}_{\tilde{t}}(k) = \mat{G}_{\tilde{t}+k} \mat{a}_{\tilde{t}}(k-1)$, $\mat{R}_{\tilde{t}}(k) = \mat{G}_{\tilde{t}+k} \mat{R}_{\tilde{t}}(k-1) \mat{G}_{\tilde{t}+k}^\top + \mat{W}_{\tilde{t}+k}$, $\mat{a}_{\tilde{t}}(0) = \mat{m}_{\tilde{t}}$, and $\mat{R}_{\tilde{t}}(0) = \mat{C}_{\tilde{t}}$, with $\mat{m}_{\tilde{t}}$ and $\mat{C}_{\tilde{t}}$ denoting the filtered mean and covariance of $\vect{\theta}_{\tilde{t}}$, respectively. This distribution is implemented in the standard \code{predict()} method for fitted dynamic objects and is available for both MCMC and LDVB fits. The fitted object supplies the filtered posterior summaries at the forecast origin, while optional future observation and evolution matrices are supplied when the required forecast design is not already stored in the fitted model. Table \ref{table:fore} summarizes the forecast inputs and optional posterior predictive draw controls. The returned
\code{exdqlmForecast} object contains the forecast-quantile summaries and state moments, with posterior-predictive draws included when requested. Forecast objects can then be plotted or evaluated against held-out observations using \code{diagnostics()}.

\begin{table}[!htb]
\centering
\TableStyle
\begin{tabularx}{\textwidth}{@{}>{\raggedright\arraybackslash}p{0.18\textwidth}>{\centering\arraybackslash}p{0.20\textwidth}>{\raggedright\arraybackslash}p{0.18\textwidth}>{\raggedright\arraybackslash}X@{}}
	\toprule
	Quantity & Mathematical symbol & \proglang{R} argument & Input/default \\
	\midrule
			Fitted dynamic model & $(\mat{m}_{\tilde{t}}, \mat{C}_{\tilde{t}})$ & \code{object} & Required fitted dynamic object from \code{exdqlmMCMC()}, \code{exdqlmLDVB()}, or legacy \code{exdqlmISVB()}. \\
	Forecast origin & $\tilde{t}$ & \code{start.t} & Required integer time index within the fitted model. \\
	Forecast horizon & $k$ & \code{k} & Required positive integer number of forecast steps. \\
		Future observation vectors & $\mat{F}_{(\tilde{t}+1):(\tilde{t}+k)}$ & \code{fFF} & Optional $q \times 1$ or $q \times k$ matrix; required with \code{fGG} when forecasting beyond the stored model matrices. \\
		Future evolution matrices & $\mat{G}_{(\tilde{t}+1):(\tilde{t}+k)}$ & \code{fGG} & Optional $q \times q$ matrix or $q \times q \times k$ array; required with \code{fFF} when forecasting beyond the stored model matrices. \\
	Credible interval mass & $1-\alpha$ & \code{cr.percent} & Optional scalar in $(0,1)$; default \code{0.95}. \\
	Forecast draws & $y^{rep}_{\tilde{t}+1:\tilde{t}+k}$ & \code{return.draws}; \code{n.samp}; \code{seed} & Optional controls; default \code{return.draws = FALSE}. When requested, \code{n.samp} sets the number of draw columns and \code{seed} makes generation reproducible. \\
	\bottomrule
\end{tabularx}
\caption{Forecast notation and principal arguments to the \code{predict()} method for fitted dynamic objects. Here $q$ denotes the state dimension.}
\label{table:fore}
\end{table}


\section{Package design and implementation}
\label{sec:design}

The package interface follows three main design principles. First, model specification is separated from estimation so that the same state-space model can be used with different inference engines. Second, engine-specific S3 classes are combined with shared family classes to provide common behavior without breaking existing code. Third, post-processing operations return explicit reusable objects, separating computation from printing, summarization, and visualization. Table 6 summarizes the resulting workflow.

Model specification functions such as \code{polytrendMod()}, \code{seasMod()}, and \code{regMod()} return objects of class \code{exdqlm}. These objects store the state-space matrices, initial-state prior, and component structure used by the fitting routines, and they can be combined with the \code{+} method to build larger dynamic models from simpler components.  This structure allows a single model object to be used with different inference engines, AL/DQLM settings, or transfer-function extensions.

Fitting routines return list-based S3 objects whose class vectors include an engine-specific first class and a shared family class. Dynamic fits keep \code{exdqlmLDVB}, \code{exdqlmMCMC}, or \code{exdqlmISVB} as their engine-specific class and also inherit from the shared \code{exdqlmFit} family. Static fits keep \code{exalStaticLDVB} or \code{exalStaticMCMC} as
their engine-specific class and inherit from \code{exalStaticFit}. The shared family classes provide common behavior while preserving the engine-specific class names used by existing code. Fitted objects can be inspected with \code{print()} and \code{summary()}, and dynamic fits support \code{plot()} for fitted quantiles, component contributions, and state summaries.

Post-processing uses S3 methods when an operation is defined on an existing fitted or forecast object. Thus, \code{predict()} and \code{diagnostics()} dispatch on the class of their input, while \code{print()}, \code{summary()}, and \code{plot()} operate on the objects they return. Named constructors are used for operations that combine several objects or draw matrices rather than acting on a single object. In particular, \code{quantileSynthesis()} remains a named function because it combines posterior-predictive information from several separately fitted quantile levels. Named helper functions are retained where needed for backward compatibility, but the method-based interface is the preferred workflow for new code.

For dynamic fit objects in the \code{exdqlmFit} family, \code{predict()} returns an \code{exdqlmForecast} object and \code{diagnostics()} returns an \code{exdqlmDiagnostic} object. Forecast objects store forecast quantiles and, when requested, posterior predictive forecast draws. They can be inspected with \code{print()}, summarized with \code{summary()}, plotted with \code{plot()}, and scored against held-out observations with \code{diagnostics()}. The latter method returns an \code{exdqlmForecastDiagnostic} object. Dynamic diagnostic objects contain calibration and predictive diagnostics for the fitted model described above and can be inspected with \code{print()}, summarized with \code{summary()}, and visualized with \code{plot()}.

The function \code{quantileSynthesis()} combines posterior predictive draws from separately fitted quantiles into a single synthesized posterior predictive distribution and returns an \code{exdqlmSynthesis} object. Static fitted objects in the \code{exalStaticFit} family support fitted-quantile plots through \code{plot()} and static model assessment through \code{diagnostics()}. The latter returns an  \code{exalStaticDiagnostic} object with fitted-quantile summaries, optional holdout or reference metrics, and coefficient summaries that can be printed, summarized, or plotted.

The object system also provides several extension points. Compatible user-supplied state-space specifications can be converted to \code{exdqlm} objects through \code{as.exdqlm()} and combined with existing model components through the \code{+} method. For developers, the shared fitted-object families provide a pattern for incorporating additional inference engines: an engine-specific class can also include \code{exdqlmFit} or \code{exalStaticFit} in its class vector and thereby use the common methods, with specialized methods supplied only when its behavior differs. Additional displays and post-processing procedures can similarly be implemented as S3 methods for the fitted, forecast, diagnostic, or synthesis object families.

Computationally intensive operations have \proglang{C++} implementations through \pkg{Rcpp},  with \proglang{R} implementations retained for numerical parity checks and portability. These include the exported exAL density, distribution, quantile, and random-generation utilities; selected Kalman recursions and model builders; posterior and latent-variable sampling paths; posterior predictive simulation; and MCMC backend routing. Package options control backend routing for Kalman recursions, model builders, posterior sampling, posterior predictive simulation, and MCMC routing; OpenMP-enabled paths also respect the \code{exdqlm.cpp\_threads} thread cap.

The variational routines return standardized traces for the ELBO, scale, skewness, and convergence diagnostics, with the closed-form and Laplace--delta ELBO components documented in the appendices. The examples use a fixed backend profile, recorded random-number seeds, and a documented reference environment so that run times and stochastic outputs are reproducible.
These computational controls affect implementation and performance rather than the statistical models being fitted.

\begin{table}[!htbp]
\centering
\TableStyle
\begin{tabularx}{\textwidth}{
  @{}
  >{\raggedright\arraybackslash}p{0.25\textwidth}
  >{\raggedright\arraybackslash}p{0.25\textwidth}
  >{\raggedright\arraybackslash}X
  @{}
}
\toprule
Function or method & Returned class or output & Role in the workflow \\
\midrule

\multicolumn{3}{@{}l@{}}{
  \rule{0pt}{2.2ex}\textit{Dynamic model construction}
} \\
\rule{0pt}{2.2ex}\code{polytrendMod()}
  & \code{exdqlm}
  & Create a polynomial-trend component.\\
\code{seasMod()}
  & \code{exdqlm}
  & Create a Fourier-form seasonal or periodic component.\\
\code{regMod()}
  & \code{exdqlm}
  & Create a regression component from covariates.\\
\code{as.exdqlm()}
  & \code{exdqlm}
  & Coerce a compatible list or time-invariant \code{dlm} object.\\

\addlinespace[1pt]
\multicolumn{3}{@{}l@{}}{
  \rule{0pt}{2.2ex}\textit{Dynamic model fitting}
  -- Input \code{exdqlm} objects
} \\
\rule{0pt}{2.2ex}\code{exdqlmMCMC()}
  & \code{exdqlmMCMC}, \code{exdqlmFit}
  & Fit a dynamic exDQLM or DQLM using MCMC.\\
\code{exdqlmLDVB()}
  & \code{exdqlmLDVB}, \code{exdqlmFit}
  & Fit a dynamic exDQLM or DQLM using LDVB.\\
\code{exdqlmTransferLDVB()}
  & \code{exdqlmLDVB}, \code{exdqlmFit}
  & Fit a transfer-function exDQLM using LDVB.\\
\code{exdqlmTransferMCMC()}
  & \code{exdqlmMCMC}, \code{exdqlmFit}
  & Fit a transfer-function exDQLM using MCMC.\\

\addlinespace[1pt]
\multicolumn{3}{@{}l@{}}{
  \rule{0pt}{2.2ex}\textit{Static model fitting}
  -- Input design matrix and response
} \\
\rule{0pt}{2.2ex}\code{exalStaticLDVB()}
  & \code{exalStaticLDVB}, \code{exalStaticFit}
  & Fit static exAL or AL regression using LDVB.\\
\code{exalStaticMCMC()}
  & \code{exalStaticMCMC}, \code{exalStaticFit}
  & Fit static exAL or AL regression using MCMC.\\

\addlinespace[1pt]
\multicolumn{3}{@{}l@{}}{
  \rule{0pt}{2.2ex}\textit{Dynamic fit examination}
  -- Input \code{exdqlmFit} objects
} \\
\rule{0pt}{2.2ex}\code{print()} / \code{summary()}
  & Printed or structured summary
  & Inspect the class, engine, sample size, state dimension,
    stored draws, convergence, and run time.\\
\code{plot()}
  & Graphical display
  & Plot fitted dynamic quantiles, component contributions,
    or state summaries.\\
\code{predict()}
  & \code{exdqlmForecast}
  & Compute forecast quantiles and optional posterior-predictive draws.\\
\code{diagnostics()}
  & \code{exdqlmDiagnostic}
  & Compute calibration and predictive diagnostics;
    use \code{plot()} to visualize them.\\

\addlinespace[1pt]
\multicolumn{3}{@{}l@{}}{
  \rule{0pt}{2.2ex}\textit{Forecast examination}
  -- Input \code{exdqlmForecast} objects
} \\
\rule{0pt}{2.2ex}\code{print()} / \code{summary()} / \code{plot()}
  & Summary or graphical display
  & Inspect and visualize forecast-quantile summaries
    and posterior-predictive intervals.\\
\code{diagnostics()}
  & \code{exdqlmForecastDiagnostic}
  & Score forecasts against held-out observations.\\

\addlinespace[1pt]
\multicolumn{3}{@{}l@{}}{
  \rule{0pt}{2.2ex}\textit{Static fit examination}
  -- Input \code{exalStaticFit} objects
} \\
\rule{0pt}{2.2ex}\code{print()} / \code{summary()}
  & Printed or structured summary
  & Inspect the class, engine, sample size, design dimension,
    stored draws, and run time.\\
\code{plot()}
  & Graphical display
  & Plot fitted conditional-quantile summaries from static
    LDVB or MCMC fits.\\
\code{diagnostics()}
  & \code{exalStaticDiagnostic}
  & Compute fitted-quantile diagnostics and coefficient summaries;
    use \code{plot()} for quantile or coefficient displays.\\

\addlinespace[1pt]
\multicolumn{3}{@{}l@{}}{
  \rule{0pt}{2.2ex}\textit{Predictive synthesis}
  -- Input dynamic fit or forecast objects, or draw matrices
} \\
\rule{0pt}{2.2ex}\code{quantileSynthesis()}
  & \code{exdqlmSynthesis}
  & Synthesize posterior-predictive draws across separately fitted
    quantile levels; use \code{print()}, \code{summary()}, or
    \code{plot()} to examine the result.\\

\bottomrule
\end{tabularx}

\caption{
Core object families, workflow functions, and standard methods used in
the examples. For fitted models, the two classes shown in the second
column are elements of the object's S3 class vector: the engine-specific
class is retained for backward compatibility, while \code{exdqlmFit} and
\code{exalStaticFit} provide shared behavior for dynamic and static fits,
respectively. Post-processing operations return explicit objects that can
be printed, summarized, or plotted where a corresponding method is defined.
}
\label{table:flow}
\end{table}


\section{Examples}
 \label{sec:exs}

We demonstrate the general workflow and key features of the package through three real-data analyses and one simulation benchmark. The Lake Huron analysis illustrates dynamic model construction, MCMC inference, forecasting, and posterior-predictive synthesis across separately fitted quantiles. The sunspot analysis demonstrates model-component composition, LDVB inference, fitted-model diagnostics, discount-factor comparison, and comparison with MCMC. The Big Tree water-flow analysis compares specifications without covariates, with direct regression effects, and with transfer-function effects, including held-out forecasting. Finally, the simulation example illustrates static exAL regression with shrinkage priors and compares LDVB with MCMC.

The package also exposes distribution-level utilities for the exAL error family: \code{dexal()}, \code{pexal()}, \code{qexal()}, and \code{rexal()} evaluate the density, distribution function, quantile function, and random generator, respectively, while  \code{get\_gamma\_bounds()} returns the admissible interval for $\gamma$ at a specified \(p_0\). These utilities are useful for simulation, prior calibration, and direct checks of the exAL distribution, but are not needed for the main fitting, forecasting, diagnostic, or synthesis workflows illustrated below. We use CrI as shorthand for credible interval in figure captions and plotting labels.

We begin by loading the \pkg{exdqlm} package used for all examples.

\begin{CodeChunk}
\begin{CodeInput}
R> library("exdqlm")
\end{CodeInput}
\end{CodeChunk}


\subsection{Lake Huron}
\label{sec:ex1}

\begin{figure}[!t]
\centering
\includegraphics[width=0.9\textwidth]{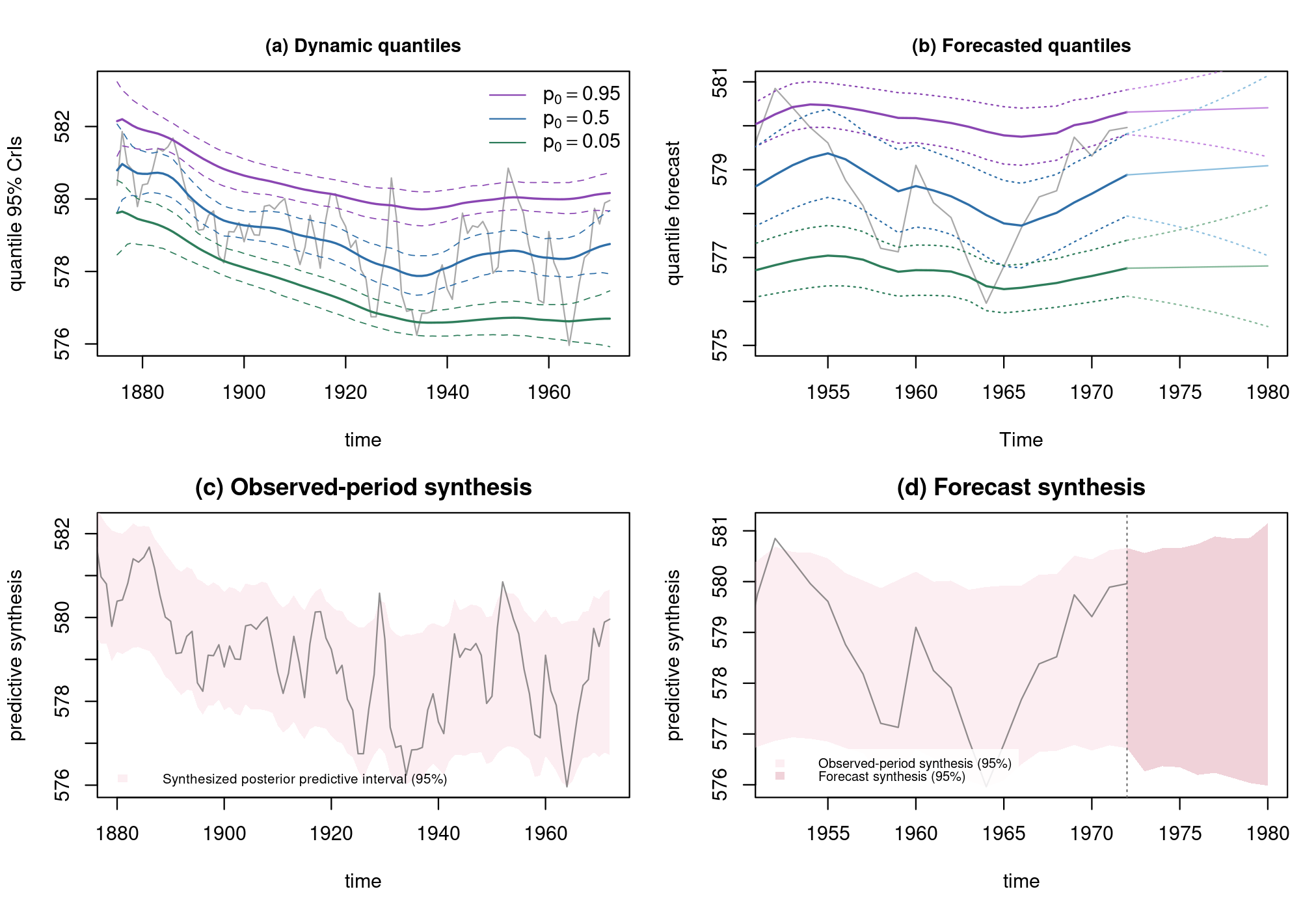}
\caption{Example 1: Lake Huron. Top-left: posterior mean estimates and 95\% CrIs of the estimated quantiles, plotted with the data (grey). Top-right: forecasted quantile estimates and 95\% credible intervals (seen after 1972), along with the filtered quantile estimates and 95\% credible intervals (seen from 1952 to 1972). Bottom-left: synthesized posterior predictive 95\% interval over the observed period. Bottom-right: observed-period synthesis near the forecast origin (lighter band) and synthesized posterior predictive 95\% interval over the eight-step forecast window (slightly darker rose band); the vertical dotted line marks the forecast origin.}
\label{fig:ex1quants}
\end{figure}

In this example, we consider the Lake Huron time series from the package \pkg{datasets}. The data are annual measurements of the level (ft) of Lake Huron from 1875 to 1972, shown in Figure \ref{fig:ex1quants}. This example shows how to build a basic state-space structure, specify a prior on $\gamma$, run the MCMC algorithm, and plot estimated quantiles and forecast distributions.

To estimate the dynamic distribution of the data, we consider three quantiles, $p_0=0.95,0.50,$ and $0.05$. We begin by creating the state-space model structure and the prior parameters (i.e. $\mat{F}_t,\mat{G}_t, \mat{m}_0$, and $\mat{C}_0$ discussed in Section \ref{sec:exdqlm}). We choose to model the quantiles with a second order polynomial trend, which we construct with the \code{polytrendMod()} function. The initial level prior is centered at 579 ft, a rounded domain-scale value chosen before fitting rather than the sample mean, and the initial slope prior is centered at zero.
\begin{CodeChunk}
\begin{CodeInput}
R> model = polytrendMod(order = 2, m0 = c(579, 0), C0 = 10 * diag(2))
\end{CodeInput}
\end{CodeChunk}

As an illustration we fix a single discount factor of $0.9$ to define the evolution covariance matrix $\mat{W}_t$.
 Discount factors can also be selected using in-sample diagnostic criteria, as illustrated in Section \ref{sec:ex2}.
 The prior location parameters for \(\gamma\) are quantile-specific: they are set to zero for the median and shifted in opposite directions for the upper and lower tails to provide tail-specific regularization within the admissible support. In these fits, the scale parameter is 0.1 and the number of degrees of freedom is 1. These quantities specify prior information and do not determine the MCMC starting values.
 In the fits shown here we use \code{n.burn = 2000} and \code{n.mcmc = 3000}. The computation uses the backend settings described in Section \ref{sec:design}; backend and runtime provenance for the manuscript run are recorded in the replication logs.

\begin{CodeChunk}
\begin{CodeInput}
R> set.seed(20260501)
R> M95 = exdqlmMCMC(y = LakeHuron, p0 = 0.95, model = model, df = 0.9, dim.df = 2,
+                 PriorGamma = list(m_gam = -1, s_gam = 0.1, df_gam = 1),
+                 n.burn = 2000, n.mcmc = 3000, verbose = FALSE)
R> M5 = exdqlmMCMC(y = LakeHuron, p0 = 0.05, model = model, df = 0.9, dim.df = 2,
+                 PriorGamma = list(m_gam = 1, s_gam = 0.1, df_gam = 1),
+                 n.burn = 2000, n.mcmc = 3000, verbose = FALSE)
\end{CodeInput}
\end{CodeChunk}

\begin{figure}[t]
\centering
\includegraphics[width=0.82\textwidth]{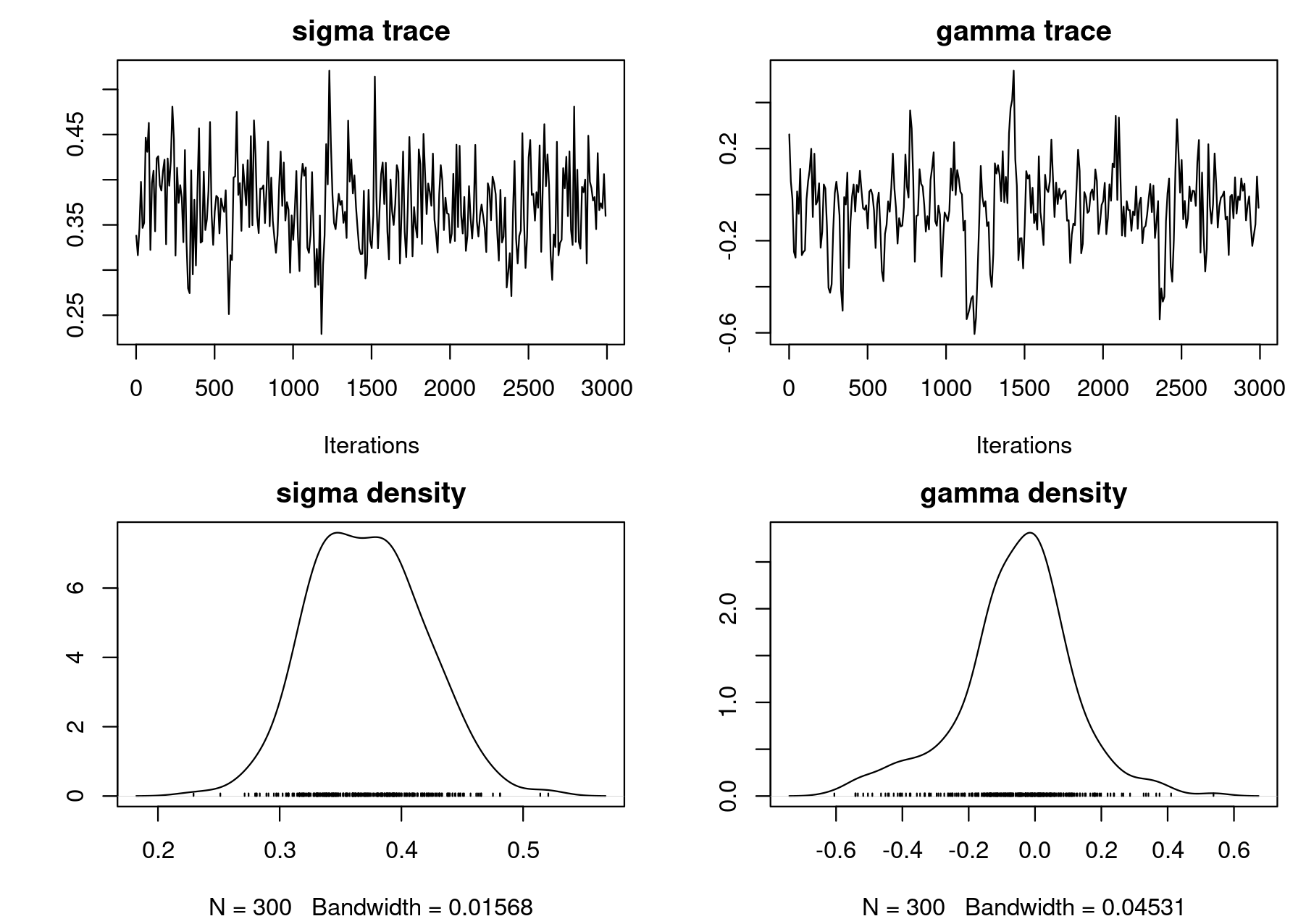}
\caption{Example 1: Lake Huron. Trace and density plots from the median run with 7000 burn-in iterations and 3000 retained draws. For readability, every 10th retained draw is used in the plotting step. The top row shows the posterior trace plots for \(\sigma\) and \(\gamma\), and the bottom row shows the corresponding posterior densities. The displayed paths are representative MCMC traces from the recorded run profile; exact trace paths may differ across reruns while retaining comparable posterior summaries.}
\label{fig:ex1mcmc}
\end{figure}

By default, \code{exdqlmMCMC()} uses the package's LDVB routine, \code{exdqlmLDVB()}, to warm-start the chain. This initialization choice is separate from the subsequent MCMC kernel. Because the data are relatively symmetric, we do not expect the skewness parameter \(\gamma\) to be far from zero at \(p_0 = 0.5\). For the diagnostic display, we run the unrestricted median model separately as \code{M50.trace} with a longer burn-in. Consistent with the symmetry of the data, the posterior mean of \(\gamma\) is \(-0.063\) with a 95\% credible interval \((-0.484, 0.282)\), while the posterior mean of \(\sigma\) is \(0.372\) with a 95\% credible interval \((0.281, 0.472)\). All retained samples in the output of \code{exdqlmMCMC()} are \code{mcmc} objects, so we can use the package \pkg{coda} \citep{coda} to examine the trace and density plots of both \(\sigma\) and \(\gamma\). For readability, we plot every 10th retained draw in Figure \ref{fig:ex1mcmc}.

\begin{CodeChunk}
\begin{CodeInput}
R> set.seed(20260620)
R> M50.trace = exdqlmMCMC(y = LakeHuron, p0 = 0.50, model = model, df = 0.9, dim.df = 2,
+                         PriorGamma = list(m_gam = 0, s_gam = 0.1, df_gam = 1),
+                         n.burn = 7000, n.mcmc = 3000,verbose = FALSE)
R> par(mfcol = c(2, 2), mar = c(4.1, 4.1, 2.1, 1.0))
R> keep.idx = seq(1, length(M50.trace$samp.sigma), by = 10)
R> sigma.trace = coda::mcmc(M50.trace$samp.sigma[keep.idx], thin = 10)
R> gamma.trace = coda::mcmc(M50.trace$samp.gamma[keep.idx], thin = 10)
R> coda::traceplot(sigma.trace, main = "sigma trace")
R> coda::densplot(sigma.trace, main = "sigma density")
R> coda::traceplot(gamma.trace, main = "gamma trace")
R> coda::densplot(gamma.trace, main = "gamma density")
\end{CodeInput}
\end{CodeChunk}

The unrestricted median run is used only to assess whether the additional skewness parameter is needed at \(p_0=0.50\). We use a longer burn-in for this diagnostic run because \(\gamma\) is sampled in that fit. After the posterior for \(\gamma\) is found to be concentrated near zero, the median model used in the quantile, forecast, and synthesis panels is the fixed-\(\gamma\) DQLM \code{M50.dqlm}. Because \(\gamma\) is fixed in this model, the skewness update is removed from the MCMC kernel and we use the same 2000 burn-in iterations and 3000 retained draws used for the other main fits. We re-run the model with a point-mass prior on $\gamma$ at 0 using the settings \code{gam.init = 0} and \code{fix.gamma = TRUE}, which is equivalent to the setting \code{dqlm.ind = TRUE}. This reduces the model to the DQLM mentioned in Section \ref{sec:special}.

\begin{CodeChunk}
\begin{CodeInput}
R> M50.dqlm = exdqlmMCMC(y = LakeHuron, p0 = 0.50, model = model, df = 0.9, dim.df = 2,
+             gam.init = 0, fix.gamma = TRUE, n.burn = 2000, n.mcmc = 3000, verbose = FALSE)
\end{CodeInput}
\end{CodeChunk}

The results are \code{exdqlmMCMC} objects that can be used for analysis and forecasting. First we examine the estimated quantiles with the plot method for \code{exdqlmMCMC} objects, seen in the top-left panel of Figure \ref{fig:ex1quants}. This generates a plot of the data with posterior mean estimates and 95\% CrIs of the dynamic quantile. Subsequent quantiles (i.e. $p_0 = 0.50, 0.05$ in this example) can be added to the same axes by calling \code{plot()} on the fitted objects with \code{add = TRUE}.

\begin{CodeChunk}
\begin{CodeInput}
R> par(mfrow = c(2, 2), mar = c(4.4, 4.1, 2.2, 1.2), oma = c(0, 0, 0.8, 0))
R> plot(M95); title("(a) Dynamic quantiles")
R> plot(M50.dqlm, add = TRUE, col = "blue")
R> plot(M5, add = TRUE, col = "forest green")
R> legend("topright", lty = 1, bty = "n", col = c("purple", "blue", "forest green"),
+        legend = c(expression('p'[0]*'=0.95'), expression('p'[0]*'=0.50'),
+                   expression('p'[0]*'=0.05')))
\end{CodeInput}
\end{CodeChunk}

Next, we forecast the $k=8$ step-ahead distributions starting in 1972. To do this, we first create the observational vector (\code{fFF}) and evolution matrix (\code{fGG}) that will be used in the forecast updates. In this case both are non-time-varying and identical to those used to estimate the quantile.
\begin{CodeChunk}
\begin{CodeInput}
R> fFF = model$FF
R> fGG =  model$GG
\end{CodeInput}
\end{CodeChunk}

We plot the time series data in a narrower range for closer examination. Forecast objects can be created with \code{predict()} on a fitted dynamic model.
Here we create the forecast objects first and then call \code{plot()} with \code{add = TRUE}. Notice we start the forecast at the last time index of the data, i.e. \code{start.t = length(LakeHuron)}.

\begin{CodeChunk}
\begin{CodeInput}
R> plot(LakeHuron,  xlim = c(1952, 1980), ylim = c(575, 582), col = "dark grey",
+           main = "(b) Forecasted quantiles", ylab = "forecast 95% CrIs")
R> fc95 = predict(M95, start.t = length(LakeHuron), k = 8,
+              fFF = fFF, fGG = fGG, return.draws = TRUE)
R> fc50 = predict(M50.dqlm, start.t = length(LakeHuron), k = 8,
+                fFF = fFF, fGG = fGG, return.draws = TRUE)
R> fc05 = predict(M5, start.t = length(LakeHuron), k = 8,
+                fFF = fFF, fGG = fGG, return.draws = TRUE)
R> plot(fc95, add = TRUE)
R> plot(fc50, add = TRUE, cols = c("blue", "light blue"))
R> plot(fc05, add = TRUE, cols = c("forest green", "green"))
\end{CodeInput}
\end{CodeChunk}

This generates a plot of the data with the forecasted quantile estimates and 95\% credible intervals (seen after 1972), along with the filtered quantile estimates and 95\% credible intervals (before 1972), for reference. Results can be seen in the top-right panel of Figure \ref{fig:ex1quants}. The percentage in the CrIs can be modified with the parameter \code{cr.percent}.

When several quantile models are fitted separately, the package also allows their posterior predictive draws to be combined into a single coherent predictive distribution through \code{quantileSynthesis()}. Each fitted model contributes local predictive information near its target quantile, and the synthesis step interpolates across these quantile-specific draws to obtain one posterior predictive distribution. When independently fitted quantiles induce crossing, optional monotonicity corrections based on isotonic adjustment and monotone rearrangement are available \citep{barlow1972,chernozhukov2010}. The synthesis operates on posterior predictive draws indexed by quantile level, so the input fits need not use identical skewness restrictions. For the Lake Huron example, we combine exDQLM tail fits at \(p_0=0.05\) and \(0.95\) with the fixed-\(\gamma\) DQLM median fit \code{M50.dqlm}.

\begin{CodeChunk}
\begin{CodeInput}
R> syn.obs = quantileSynthesis( draws_list = list(M5, M50.dqlm, M95), p = c(0.05, 0.50, 0.95),
+               enforce_isotonic = TRUE, rearrange = TRUE, T_expected = length(LakeHuron))
\end{CodeInput}
\end{CodeChunk}

The recorded manuscript run uses the package's isotonic adjustment and monotone rearrangement; these are the default correction settings and are shown explicitly in the displayed calls. The stored synthesized quantile summaries have no remaining crossings over the observed period.

Because the argument \code{return.draws = TRUE} was used, each object stores posterior predictive forecast draws in the return value \code{samp.fore}. This allows us to apply the same synthesis step to the eight-step forecast horizon.

\begin{CodeChunk}
\begin{CodeInput}
R> syn.fore = quantileSynthesis(draws_list = list(fc05, fc50, fc95), p = c(0.05, 0.50, 0.95),
+               enforce_isotonic = TRUE, rearrange = TRUE, T_expected = 8)
\end{CodeInput}
\end{CodeChunk}

The same correction settings are used for the forecast synthesis, and the eight-step synthesized forecast summaries are also ordered across the three input quantile levels.

The objects \code{syn.obs} and \code{syn.fore} have class \code{exdqlmSynthesis}, so the corresponding \code{plot()} method can be used to display the synthesized posterior predictive intervals. The lower panels of Figure \ref{fig:ex1quants} are produced from these objects as follows. The first synthesis panel shows the observed-period synthesized predictive interval, and the second panel uses the same observed-period synthesis near the forecast origin, overlays the forecast synthesis, and shades the one-step bridge from the final observed synthesis interval to the first forecast synthesis interval.

\begin{CodeChunk}
\begin{CodeInput}
R> synth.obs.col = adjustcolor("#F7D6DE", alpha.f = 0.40)
R> synth.fore.col = adjustcolor("#D98A9B", alpha.f = 0.38)
R> plot(syn.obs, y = LakeHuron, time = as.numeric(time(LakeHuron)), xlim = c(1880, 1972),
+         ylim = c(575.75,582.25), ylab = "predictive synthesis",
+         main = "(c) Observed-period synthesis", show.median = FALSE,
+         band.col = synth.obs.col, y.col = adjustcolor("grey30", alpha.f = 0.62))
R> legend("bottomleft", legend = "Synthesized posterior predictive interval (95%)",
+           fill = synth.obs.col, border = NA, bty = "n", cex = 0.68)
R> plot(syn.obs, y = LakeHuron, time = as.numeric(time(LakeHuron)), xlim = c(1952, 1980),
+         ylim = c(575.75,581), ylab = "predictive synthesis",
+         main = "(d) Forecast synthesis", show.median = FALSE, band.col = synth.obs.col,
+         y.col = adjustcolor("grey30", alpha.f = 0.62))
R> polygon(c(1972, 1973, 1973, 1972), c(tail(syn.obs$summary$q025, 1),
+         syn.fore$summary$q025[1], syn.fore$summary$q975[1],
+         tail(syn.obs$summary$q975, 1)), col = synth.fore.col, border = NA)
R> plot(syn.fore, time = 1973:1980, add = TRUE, show.median = FALSE,
+          band.col = synth.fore.col)
R> abline(v = 1972, lty = 3)
R> legend("bottomleft", legend = c("Observed-period synthesis (95%)",
+          "Forecast synthesis (95%)"),
+          fill = c(synth.obs.col, synth.fore.col), border = NA, bty = "n", cex = 0.66)
\end{CodeInput}
\end{CodeChunk}


\subsection{Sunspots}
\label{sec:ex2}

\begin{figure}[t]
\centering
\includegraphics[width=1\textwidth]{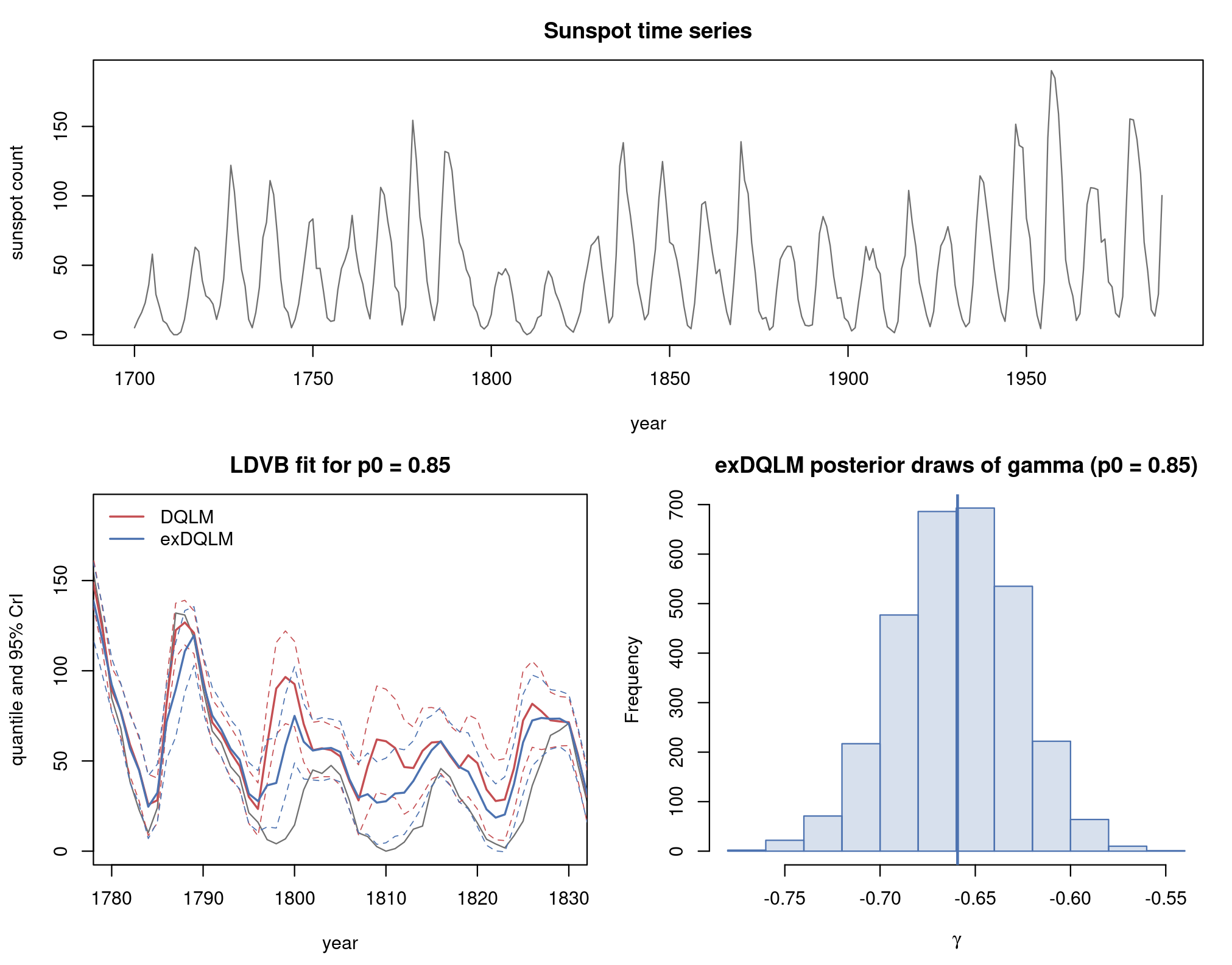}
\caption{Example 2: Sunspots. Top: The sunspot time series from 1700 to 1988. Bottom left: posterior mean estimates and 95\% CrIs of the estimated quantiles from the DQLM (red) and exDQLM (blue), plotted with the data (grey) from 1780 to 1830. Bottom right: Histogram of samples from the approximated posterior distribution of $\gamma$ under the exDQLM.}
\label{fig:ex2quant}
\end{figure}

For our next example, we use the yearly Sunspot time series from the package \pkg{datasets}. The data are yearly counts for sunspots from 1700 to 1988, shown in Figure \ref{fig:ex2quant}. In this example we show how to use the \pkg{dlm} package to create the state-space model, combine blocks of a state-space model, apply the LDVB approximation through \code{exdqlmLDVB()}, perform visual model diagnostics to compare the exDQLM with the DQLM, and use those diagnostics for discount-factor selection.

Upper-tail sunspot activity is a natural setting for illustrating dynamic quantile modeling of cyclic count data. We therefore model the \(0.85\) quantile in this example. Although sunspot activity is recorded as a count, we use the continuous exAL working likelihood to model its upper conditional quantile; the example is not intended as a discrete-count likelihood analysis. Again, we begin by building the state-space structure used to estimate the quantile. Here we choose to model the quantile with a first order polynomial trend component as well as a Fourier form seasonal component that incorporates a fundamental period every 11 observations (years), as well as some of its corresponding harmonics. This results in a dynamic model with a total of 9 state parameters, 1 for the first order polynomial component and 8 for the seasonal component. The \pkg{exdqlm} functions are compatible with time-invariant \code{dlm} objects from the \pkg{dlm} package. For example, we create the first order trend component with the function \code{dlmModPoly()} from the \pkg{dlm} package, then use \code{as.exdqlm()} to turn the output into an \code{exdqlm} object.
\begin{CodeChunk}
\begin{CodeInput}
R> dlm.trend.comp = dlm::dlmModPoly(1, m0 = 50, C0 = 2500)
R> trend.comp = as.exdqlm(dlm.trend.comp)
\end{CodeInput}
\end{CodeChunk}
The initial level prior is centered at 50 sunspots, a rounded count-scale value chosen before fitting, with prior standard deviation 50 to keep the level weakly informative on the observed count scale. Next, we construct the seasonal component with our \code{seasMod()} function. Sunspot cycles are commonly modeled with an approximately 11-year period; we include the first four Fourier harmonics in the component.
\begin{CodeChunk}
\begin{CodeInput}
R> seas.comp = seasMod(p = 11, h = 1:4, C0 = 10 * diag(8))
\end{CodeInput}
\end{CodeChunk}
To combine the two components into a single state-space structure, we add the two \code{exdqlm} objects.
\begin{CodeChunk}
\begin{CodeInput}
R> model = trend.comp + seas.comp
\end{CodeInput}
\end{CodeChunk}
Adding the two model components produces a block-composed state-space model with one trend state and eight seasonal states.

We will apply discount factors by component, i.e. a discount factor of $0.9$ for the trend component of dimension $1$ and a discount factor of $0.85$ for the seasonal component of dimension $8$. This discounting strategy can be applied with arguments \code{df = c(0.9, 0.85)} and \code{dim.df = c(1, 8)}.

We call the function \code{exdqlmLDVB()} to obtain the LDVB approximations. We use the routine with argument \code{dqlm.ind = TRUE} to estimate the DQLM, as well as with the default settings to estimate the exDQLM. In this example we set \code{sig.init = 2} as the starting value for the LDVB scale update, use \code{n.samp = 3000}, and suppress iteration output with \code{verbose = FALSE}.

\begin{CodeChunk}
\begin{CodeInput}
R> set.seed(20262601)
R> M1 = exdqlmLDVB(y = sunspot.year, p0 = 0.85, model = model, df = c(0.9, 0.85),
+            dim.df = c(1, 8), dqlm.ind = TRUE, fix.sigma = FALSE,
+            sig.init = 2, n.samp = 3000, verbose = FALSE)
R> set.seed(20262602)
R> M2 = exdqlmLDVB(y = sunspot.year, p0 = 0.85, model = model, df = c(0.9, 0.85),
+            dim.df = c(1, 8), sig.init = 2, fix.sigma = FALSE,
+            n.samp = 3000, verbose = FALSE)
\end{CodeInput}
\end{CodeChunk}

The fitted objects can be summarized with the \code{summary()} generic, which reports the model class, sample size, state dimension, discount factors, convergence iterations, and run time. For this dataset of length 289, the DQLM LDVB fit completes more quickly, while the corresponding exDQLM LDVB fit is more computationally demanding because the skewness parameter is estimated rather than fixed. The timing comparison in Table \ref{tab:ex2bench} reports the benchmark values used in the manuscript under the recorded backend profile. These runtimes are fit elapsed times stored in the returned fit objects; they are intended for within-profile comparison and should not be interpreted as machine-independent timing constants.

We can visualize the differences between the two resulting dynamic quantiles with the fitted-object \code{plot()} method, shown in the lower-left panel of Figure \ref{fig:ex2quant}. For clarity we limit that panel to years 1780 to 1830.

\begin{CodeChunk}
\begin{CodeInput}
R> plot(sunspot.year, xlim = c(1780, 1830), col = "dark grey", ylab = "quantile 95% CrIs")
R> plot(M1, add = TRUE, col = "red")
R> plot(M2, add = TRUE, col = "blue")
R> legend("topleft", lty = 1, bty = "n", col = c("red", "blue"), legend= c("DQLM", "exDQLM"))
R> title("LDVB fit for p0 = 0.85")
\end{CodeInput}
\end{CodeChunk}

\begin{figure}[t]
\centering
\includegraphics[width=1\textwidth]{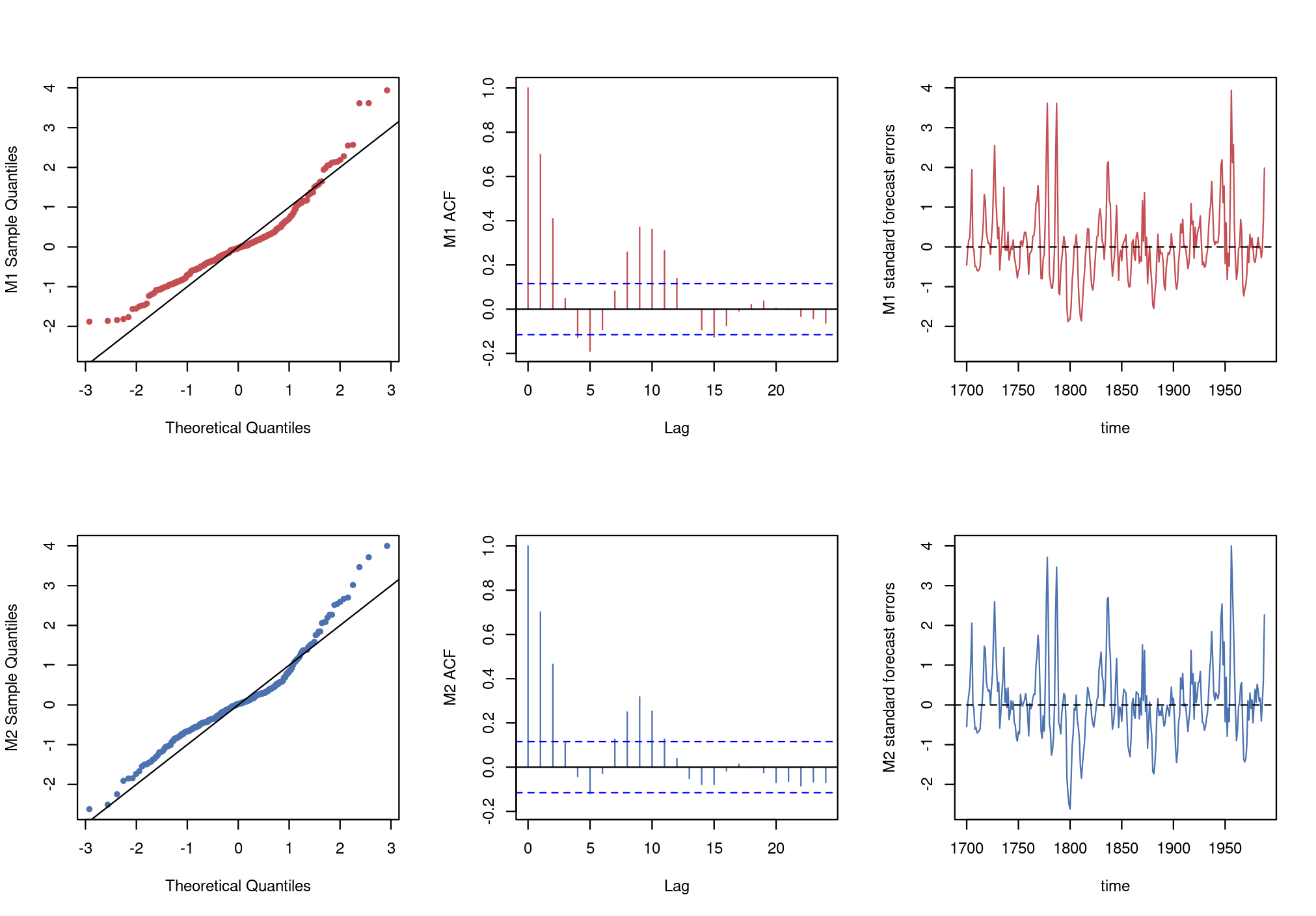}
\caption{Example 2: Sunspots. The QQ plots (left column), ACF plots of the PIT sequence (center column), and MAP one-step-ahead standardized forecast errors (right column) from the DQLM (red) and exDQLM (blue).}
\label{fig:ex2checks}
\end{figure}

To examine whether the added flexibility of the exDQLM is supported by the fitted approximation, we can also visualize the approximate posterior distribution of the skewness parameter $\gamma$ by plotting the samples from the corresponding variational distribution, also seen in Figure \ref{fig:ex2quant}. The approximate posterior mass lies away from zero, supporting the additional skewness flexibility for this upper-quantile fit.

\begin{CodeChunk}
\begin{CodeInput}
R> hist(M2$samp.gamma, xlab = expression(gamma), main = "",
+        col = adjustcolor("#4C72B0", alpha.f = 0.22), border = "#4C72B0")
R> abline(v = median(M2$samp.gamma), col = "#4C72B0", lwd = 2)
R> title("exDQLM posterior draws of gamma (p0 = 0.85)")
\end{CodeInput}
\end{CodeChunk}

To further examine the two models, we use the \code{diagnostics()} method to create an object of class \code{exdqlmDiagnostic} and then call its \code{plot()} method. The results are seen in Figure \ref{fig:ex2checks}. These diagnostic plots are based on the MAP one-step-ahead standardized forecast errors produced by the filtering recursion, together with the corresponding probability integral transform (PIT) sequence. Thus the figure summarizes the fitted diagnostic sequence rather than posterior bands for residuals.

\begin{CodeChunk}
\begin{CodeInput}
R> par(mfrow = c(2, 3))
R> diagM1M2 = diagnostics(M1, M2)
R> plot(diagM1M2, cols = c("red", "blue"))
\end{CodeInput}
\end{CodeChunk}

The \code{diagnostics()} method can also be used for a coarse in-sample comparison of candidate model specifications. Here we examine whether the discount factor used for the seasonal component is reasonable under the reported diagnostic criteria. We create a grid of possible discount factors, run the LDVB approximation for each candidate model, and compare the resulting CRPS and KL values. Because these criteria are computed using the observations employed to fit each model, this exercise is a diagnostic screening rather than an out-of-sample predictive validation. We use CRPS as the primary comparison criterion and KL divergence as a complementary calibration check; the selected discount-factor specification is the row with the smallest CRPS.

\begin{CodeChunk}
\begin{CodeInput}
R> possible.dfs = cbind(0.9, seq(0.85, 1, 0.05))
R> possible.dfs
\end{CodeInput}
\begin{CodeOutput}
     [,1] [,2]
[1,]  0.9 0.85
[2,]  0.9 0.90
[3,]  0.9 0.95
[4,]  0.9 1.00
\end{CodeOutput}
\begin{CodeInput}
R> metrics <- matrix(NA_real_, nrow(possible.dfs), 2, dimnames = list(NULL, c("CRPS", "KL")))
R> for (i in 1:nrow(possible.dfs)) {
+   set.seed(20262700 + i)
+   temp.M2 = exdqlmLDVB(y = sunspot.year, p0 = 0.85, model = model, df = possible.dfs[i, ],
+                  dim.df = c(1, 8), sig.init = 2, fix.sigma = FALSE,
+                  n.samp = 3000, verbose = FALSE)
+   temp.check = diagnostics(temp.M2)
+   metrics[i, ] = c(temp.check$m1.CRPS, temp.check$m1.KL)
+ }
R> df.scan = data.frame(possible.dfs, CRPS = metrics[, "CRPS"], KL = metrics[, "KL"])
R> round(df.scan, 3)
\end{CodeInput}
\begin{CodeOutput}
   X1   X2   CRPS    KL
1 0.9 0.85  6.235 0.120
2 0.9 0.90  8.261 0.145
3 0.9 0.95 12.458 0.173
4 0.9 1.00 20.233 0.256
\end{CodeOutput}
\begin{CodeInput}
R> possible.dfs[which.min(metrics[, "CRPS"]), ]
\end{CodeInput}
\begin{CodeOutput}
[1] 0.90 0.85
\end{CodeOutput}
\end{CodeChunk}
On this coarse grid, the CRPS and KL criteria agree and both select the seasonal discount factor \(0.85\), so the choice used above is retained.

To compare the fast variational approximation with posterior simulation, we fit matched DQLM and exDQLM specifications with \code{exdqlmMCMC()}. The MCMC fits use the same state-space model, quantile level, and discount factors as the LDVB fits, with 2000 burn-in iterations and 3000 retained draws.

\begin{CodeChunk}
\begin{CodeInput}
R> set.seed(20262801)
R> M1mcmc = exdqlmMCMC(y = sunspot.year, p0 = 0.85, model = model, df = c(0.9, 0.85),
+                dim.df = c(1, 8), n.burn = 2000, n.mcmc = 3000, verbose = FALSE,
+                dqlm.ind = TRUE, fix.sigma = FALSE)
R> set.seed(20262802)
R> M2mcmc = exdqlmMCMC(y = sunspot.year, p0 = 0.85, model = model, df = c(0.9, 0.85),
+                dim.df = c(1, 8), n.burn = 2000, n.mcmc = 3000, verbose = FALSE,
+                fix.sigma = FALSE)
\end{CodeInput}
\end{CodeChunk}

\begin{table}[t!]
\centering
\TableStyle
\begin{tabular}{@{}llrrrr@{}}
 \toprule
 model & method & \shortstack{runtime\\(s)} & KL & CRPS & PPLC \\
 \midrule
 \tworowcell{DQLM} & LDVB & \textbf{26.70} & \textbf{0.133} & 8.517 & 3843.1 \\
 & MCMC & 364.54 & 0.167 & \textbf{7.608} & \textbf{2976.8} \\
 \midrule
 \tworowcell{exDQLM} & LDVB & \textbf{134.43} & \textbf{0.120} & 6.263 & 2334.8 \\
 & MCMC & 423.47 & 0.195 & \textbf{5.111} & \textbf{2190.8} \\
 \bottomrule
\end{tabular}
\caption{Example 2: representative dynamic benchmark for the DQLM and exDQLM fits at \(p_0 = 0.85\). Each row reports fit runtime together with the deterministic KL normality diagnostic, the mean CRPS from posterior predictive empirical quantiles, and the posterior predictive loss criterion (PPLC). The LDVB rows use \code{n.samp = 3000}. The MCMC rows use 2000 burn-in iterations and 3000 retained draws. Runtimes are fit elapsed times stored in the returned fit objects and exclude diagnostic calculations, plotting, table construction, and manuscript rendering. They were measured under benchmark profile B on an Intel Xeon E5-4650 CPU with \proglang{R} 4.6.0, \pkg{exdqlm} version 1.1.0, the C++ MCMC backend in \code{"fast"} mode, and \code{exdqlm.cpp\_threads = 1L}. Because the columns are the comparison metrics, bold entries indicate the smaller value within each model block; displayed ties are left unbolded.}
\label{tab:ex2bench}
\end{table}

Table \ref{tab:ex2bench} gives a compact runtime-and-diagnostic comparison. Within each model class, LDVB is substantially faster. For both DQLM and exDQLM, LDVB gives the smaller KL value, while MCMC gives smaller CRPS and PPLC values. Comparing model classes within each inferential method, the exDQLM improves the LDVB diagnostics relative to the DQLM, and under MCMC it improves the predictive-score criteria CRPS and PPLC. The comparison shows why runtime is reported together with predictive criteria rather than by itself.


\subsection{Big Tree water flow}
\label{sec:ex3}

For the third example, we use observed average monthly water flow (cubic feet
per second) at the Big Tree gauge of the San Lorenzo River in Santa Cruz
County, CA \citep{btflow}. The \pkg{exdqlm} version 1.1.0 used for this
article includes these measurements as the monthly time series \code{BTflow},
which extends through March 2026. Here we use the complete overlap between
\code{BTflow} and the package data frame \code{climateIndices},
giving 432 monthly observations from January
1987 through December 2022. The final 18 months, July 2021 through December
2022, are not used for fitting and are shown later in a conditional forecast
display. This example illustrates the transfer-function workflow: building a
baseline dynamic quantile model, adding external covariates through the
transfer-function state augmentation, examining fitted components, and
forecasting conditionally on future covariate values.

\begin{figure}[t]
\centering
\includegraphics[width=1\textwidth]{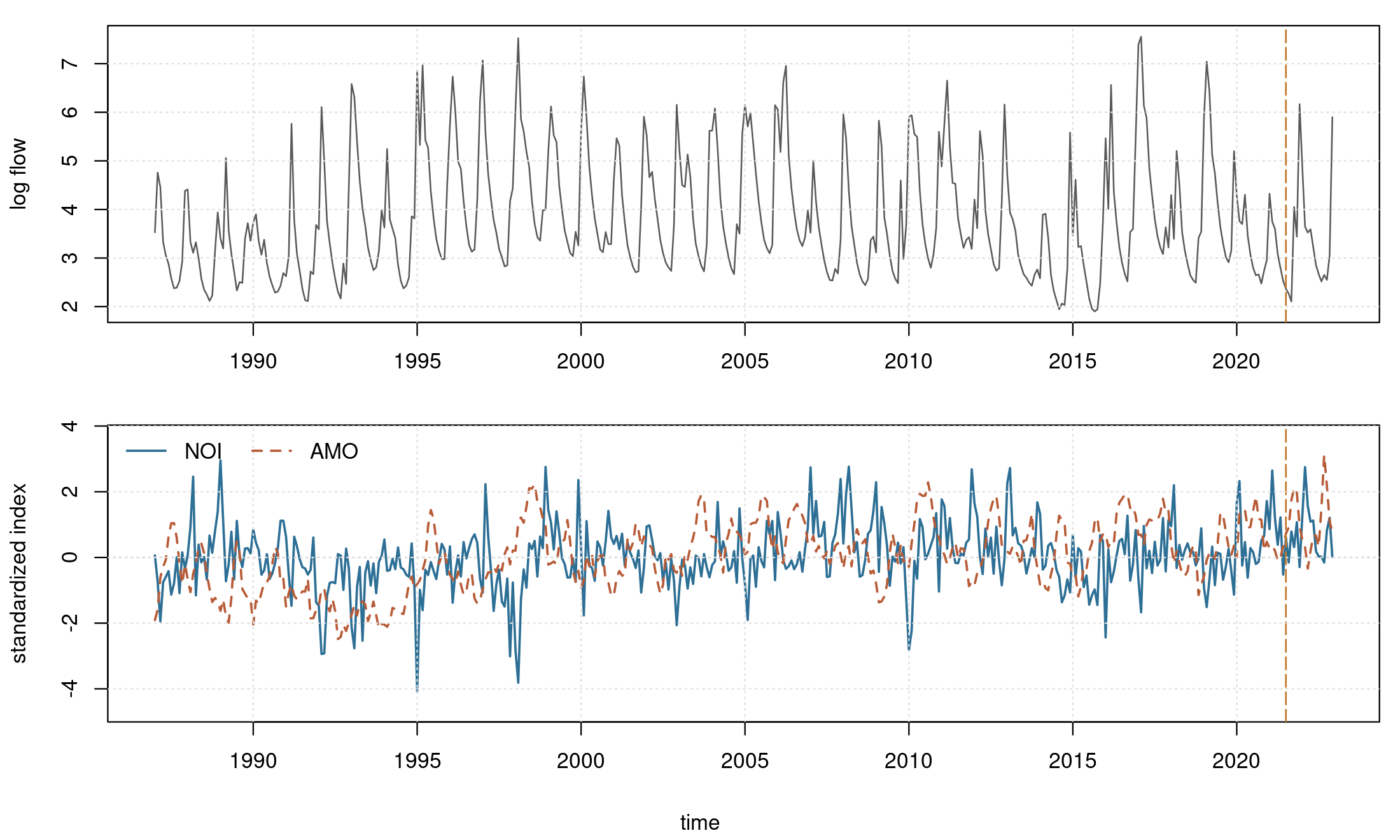}
\caption{Example 3: Big Tree water flow. Top: observed average monthly water
flow at the Big Tree gauge of the San Lorenzo River in Santa Cruz County, CA,
plotted on the log scale. Bottom: standardized monthly values of the Northern
Oscillation Index (NOI) and Atlantic Multidecadal Oscillation (AMO) index. The
displayed modeling window spans January 1987 through December 2022, and the
vertical dashed line marks the beginning of the held-out forecast window in
July 2021.}
\label{fig:ex3data}
\end{figure}

We focus on the lower-tail quantile \(p_0=0.15\), which represents lower
monthly-flow behavior on the log scale. As transfer-function inputs, we use two
monthly climate indices from \code{climateIndices}: the Northern Oscillation
Index (NOI) and the Atlantic Multidecadal Oscillation (AMO) index. These indices
come from the NOAA Physical Sciences Laboratory climate-index collection
\citep{noaaClimateIndices} and provide a compact setting for illustrating
external-input effects. The covariates are standardized using the training
window only. In the forecast display, NOI and AMO values over the held-out
months are treated as known future inputs; the display is therefore conditional
on those covariates rather than an operational forecast of the climate indices.
The complete overlap used here has 432 monthly observations. The first 414
observations, January 1987 through June 2021, form the training period; the final
18 observations, July 2021 through December 2022, form the forecast holdout. The
following chunk shows the essential data alignment. The replication script adds
date and completeness checks, uses the same training-window scaling, and writes
the aligned modeling data to
\code{analysis/manuscript/outputs/tables/ex3\_model\_dataset.csv}.

\begin{CodeChunk}
\begin{CodeInput}
R> data("BTflow", package = "exdqlm")
R> data("climateIndices", package = "exdqlm")
R> ex3.dates = seq(as.Date("1987-01-01"), by = "month", length.out = length(BTflow))
R> flow.df = data.frame(date = ex3.dates, flow = as.numeric(BTflow))
R> ex3.df = merge(flow.df, climateIndices[, c("date", "noi", "amo")], by = "date")
R> ex3.df = ex3.df[1:432, ]
R> y.fit = ts(log(ex3.df$flow), start = c(1987, 1), frequency = 12)
R> y.train = window(y.fit, end = c(2021, 6))
R> y.holdout = window(y.fit, start = c(2021, 7))
R> X.raw = as.matrix(ex3.df[, c("noi", "amo")])
R> X.train = scale(X.raw[1:414, ])
R> X.holdout = scale(X.raw[415:432, ], center = attr(X.train, "scaled:center"),
+                   scale = attr(X.train, "scaled:scale"))
\end{CodeInput}
\end{CodeChunk}

The baseline state-space model contains a first-order polynomial trend and a Fourier-form seasonal component with period 12. The seasonal block includes \(h = 1\), \(h = 2\), and \(h \approx 0.147\), corresponding to annual, semiannual, and approximately 7-year cycles on the monthly scale.
The low-frequency harmonic \(h=0.1469118636\) follows the specification developed in \citet{me}; see that work for its motivation and selection.
\begin{CodeChunk}
\begin{CodeInput}
R> trend.comp = polytrendMod(1, m0 = log(50), C0 = 1)
R> seas.comp = seasMod(p = 12, h = c(1, 2, 0.1469118636), C0 = diag(1, 6))
R> model = trend.comp + seas.comp
\end{CodeInput}
\end{CodeChunk}
The initial log-flow level prior is centered at \(\log(50)\), corresponding to
a fixed 50 cubic feet per second reference value, with prior standard deviation
1 on the log scale. This gives a broad domain-scale prior without using the
training response to center the state prior. The standardized climate coefficients are assigned zero-centered normal priors
with variance 1. On the log-flow scale this is intentionally weakly informative:
it allows either climate index to shift the lower-tail quantile materially while
still regularizing the example around no climate effect.

We compare three models built from the same trend and seasonal baseline. The
first, \code{M0}, contains no climate covariates. The second, \code{MREG}, adds
NOI and AMO as direct contemporaneous regressors using \code{regMod()}. The
third, \code{MTF}, adds the same two covariates through the transfer-function
structure from Section \ref{sec:tf}. For all three models, the trend and
seasonal discount factors are set to 0.99. To keep the climate-effect
comparison simple, the direct-regression coefficients in \code{MREG} and the
instantaneous transfer coefficients \(\psi_t\) in \code{MTF} use discount
factor 1, so these coefficient states are static over time. The transfer
accumulated-effect state \(\zeta_t\) retains discount factor 0.99.

The transfer rate \(\lambda\) is selected using the training period only. Guided
by an initial coarse scan, the replication script fits the transfer model over
\(\lambda\in\{0.70,0.75,0.80,0.85,0.90,0.95,0.99\}\) and chooses the value with
the smallest posterior predictive loss criterion (PPLC) from
\code{diagnostics()}. The held-out months are not used in this selection
step. The following code shows the selection pattern used by the replication
script. The option \code{exdqlm.max\_iter} sets the LDVB iteration cap used in
the replication run, so the displayed code records the same convergence control
as the executable script.

\begin{CodeChunk}
\begin{CodeInput}
R> lambda.grid = c(0.70, 0.75, 0.80, 0.85, 0.90, 0.95, 0.99)
R> old.opt = options(exdqlm.max_iter = 600L)
R> pplc.grid = rep(NA_real_, length(lambda.grid))
R> for (i in seq_along(lambda.grid)) {
+   set.seed(20264001 + i)
+   temp.MTF = exdqlmTransferLDVB(y = y.train, p0 = 0.15, model = model, df = c(0.99, 0.99),
+                 dim.df = c(1, 6), X = X.train, tf.df = c(0.99, 1), lam = lambda.grid[i],
+                 tf.m0 = rep(0, 3), tf.C0 = diag(c(0.1, 1, 1), 3), sig.init = 0.1,
+                 gam.init = -0.1, n.samp = 1000, tol = 0.05, verbose = FALSE)
+   temp.diag = diagnostics(temp.MTF)
+   pplc.grid[i] = temp.diag$m1.pplc
+   }
R> lambda.star = lambda.grid[which.min(pplc.grid)]
R> lambda.star
\end{CodeInput}
\begin{CodeOutput}
[1] 0.85
\end{CodeOutput}
\end{CodeChunk}

The final models are fit on January 1987 through June 2021. We use the LDVB implementation for all three models; the corresponding MCMC implementation is available when posterior simulation is desired.

\begin{figure}[!t]
\centering
\includegraphics[width=0.92\textwidth]{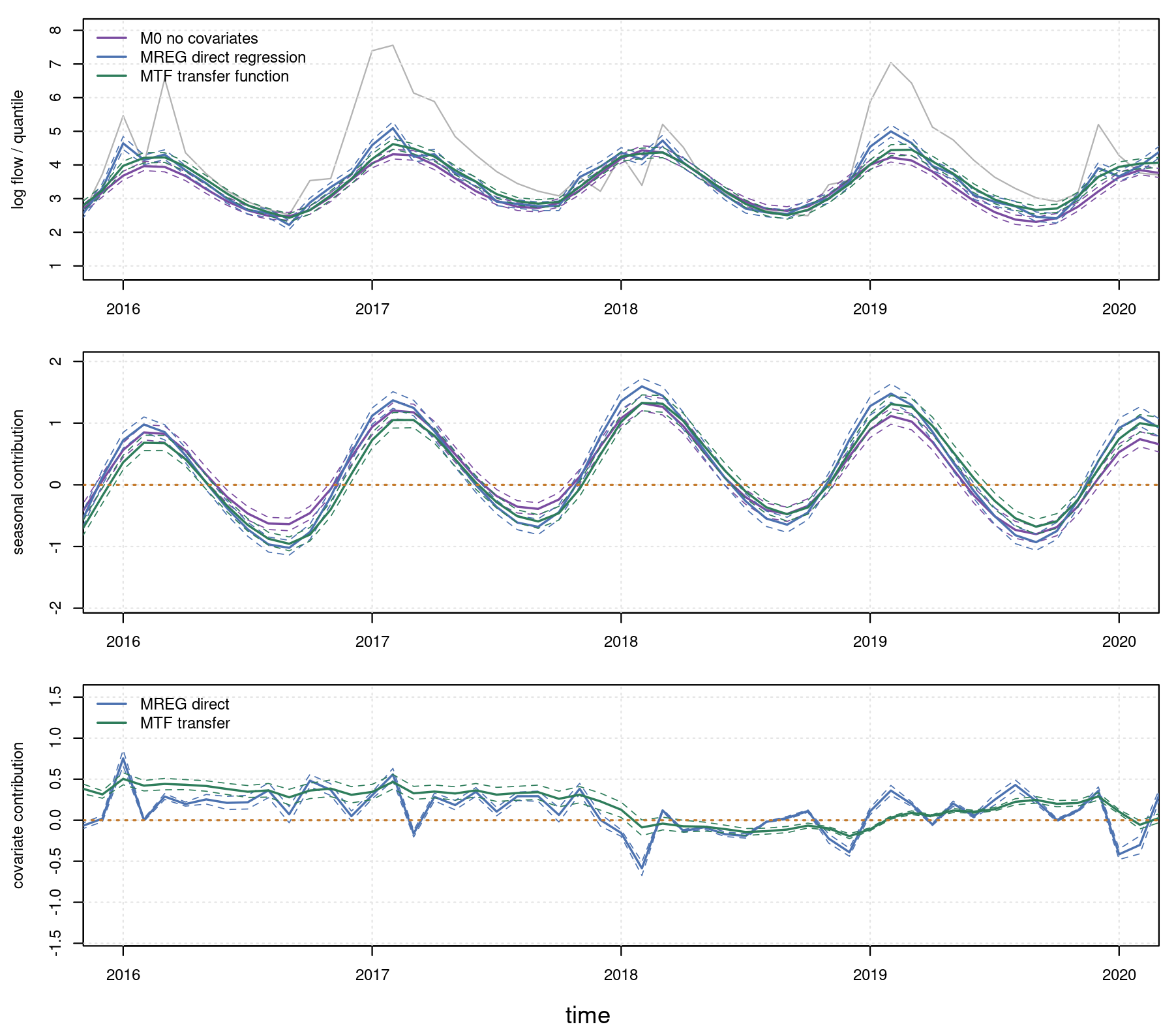}
\caption{Example 3: Big Tree water flow. Top: posterior mean estimates and 95\% CrIs of the fitted \(0.15\) quantile from the no-covariate baseline \code{M0} (purple), direct-regression model \code{MREG} (blue), and transfer-function model \code{MTF} (green), plotted with the data (grey) from 2016 to 2020. Middle: posterior mean estimates and 95\% CrIs of the combined seasonal components implied by the annual, semiannual, and low-frequency harmonics. Bottom: posterior mean estimates and 95\% CrIs of the direct-regression contribution in \code{MREG} and the transfer-function contribution in \code{MTF}. The horizontal orange dotted lines are at zero for reference.}
\label{fig:ex3quant}
\end{figure}

\begin{CodeChunk}
\begin{CodeInput}
R> set.seed(20264101)
R> M0 = exdqlmLDVB(y = y.train, p0 = 0.15, model = model, df = c(0.99, 0.99),
+                dim.df = c(1, 6), sig.init = 0.1, gam.init = -0.1,
+                n.samp = 1000, tol = 0.05, verbose = FALSE)
R> reg.comp = regMod(X.train, m0 = rep(0, 2), C0 = diag(1, 2))
R> set.seed(20264301)
R> MREG = exdqlmLDVB(y = y.train, p0 = 0.15, model = model + reg.comp, df = c(0.99, 0.99, 1),
+                dim.df = c(1, 6, 2), sig.init = 0.1, gam.init = -0.1,
+                n.samp = 1000, tol = 0.05, verbose = FALSE)
R> set.seed(20264201)
R> MTF = exdqlmTransferLDVB(y = y.train, p0 = 0.15, model = model, df = c(0.99, 0.99),
+                dim.df = c(1, 6), X = X.train, tf.df = c(0.99, 1), lam = lambda.star,
+                tf.m0 = rep(0, 3), tf.C0 = diag(c(0.1, 1, 1), 3), sig.init = 0.1,
+                gam.init = -0.1, n.samp = 1000, tol = 0.05, verbose = FALSE)
R> options(old.opt)
\end{CodeInput}
\end{CodeChunk}

The upper panel of Figure \ref{fig:ex3quant} compares the fitted \(0.15\)
quantiles from the three training fits over 2016--2020. We first plot the data, then call \code{plot()} with \code{add = TRUE} on the three fitted objects to draw the model fits on common axes.

\begin{CodeChunk}
\begin{CodeInput}
R> par(mfrow = c(3, 1), mar = c(2.8, 4.4, 1.0, 0.9), oma = c(1.8, 0, 0, 0))
R> plot(y.train, col = "grey", ylim = c(1, 8), xlim = c(2016, 2020),
+                ylab = "log flow / quantile")
R> grid(col = "grey90")
R> plot(M0, add = TRUE)
R> plot(MREG, add = TRUE, col = "steelblue")
R> plot(MTF, add = TRUE, col = "forest green")
R> legend("topleft", legend = c("M0 no covariates", "MREG direct regression",
+                "MTF transfer function"), col = c("purple", "steelblue", "forest green"),
+                lty = 1, lwd = 1.5, bty = "n")
\end{CodeInput}
\end{CodeChunk}

\begin{figure}[!t]
\centering
\includegraphics[width=0.9\textwidth]{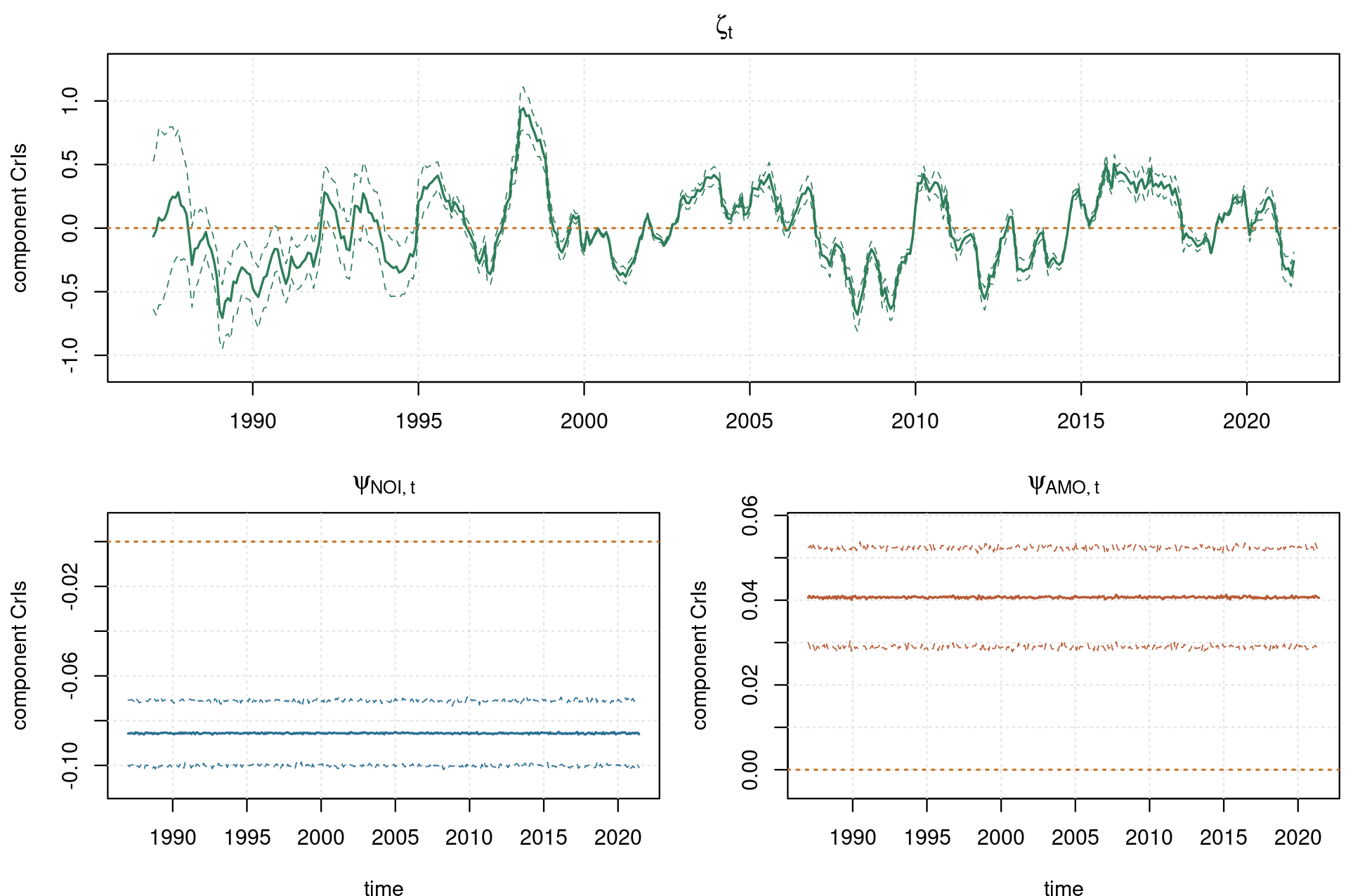}
\caption{Example 3: Big Tree water flow. Posterior mean estimates and 95\% CrIs of the transfer-function states in \code{MTF}. The top panel shows the accumulated transfer effect \(\zeta_t\). The bottom panels show the instantaneous transfer coefficients \(\psi_{\mathrm{NOI},t}\) and \(\psi_{\mathrm{AMO},t}\), which are static here because their discount factors are set to 1. Horizontal dotted orange lines mark zero.}
\label{fig:ex3tftheta}
\end{figure}

The fitted lower-tail quantile paths are similar because the trend and seasonal components explain
much of the broad annual structure. The component panels of Figure \ref{fig:ex3quant} clarify the model
differences: the middle panel shows that the seasonal contribution is similar
across models, while the lower panel separates the contemporaneous climate-index
contribution in \code{MREG} from the accumulated transfer contribution in
\code{MTF}. Component and state views are available from the fitted-object
\code{plot()} method by specifying the parameter \code{type = "component"} or \code{type = "state"}, respectively. We
begin with the combined seasonal contribution. Because the seasonal block
contains three harmonics, this contribution is formed from the second through
seventh elements of the state vector. The middle panel of Figure
\ref{fig:ex3quant} shows the resulting seasonal estimates for the three models
over the same 2016--2020 window.

\begin{CodeChunk}
\begin{CodeInput}
R> plot(NA, ylim = c(-2, 2), xlim = c(2016, 2020), ylab = "seasonal components")
R> grid(col = "grey90")
R> plot(M0, type = "component", index = 2:7, add = TRUE)
R> plot(MREG, type = "component", index = 2:7, add = TRUE, col = "steelblue")
R> plot(MTF, type = "component", index = 2:7, add = TRUE, col = "forest green")
R> abline(h = 0, col = "orange", lty = 3, lwd = 1.4)
\end{CodeInput}
\end{CodeChunk}
The lower panel of Figure \ref{fig:ex3quant} displays the climate-index
contributions induced by the standardized NOI and AMO covariates. For
\code{MREG}, indices 8 and 9 are the direct-regression coefficient states. For
\code{MTF}, index 8 is the accumulated transfer effect \(\zeta_t\). This
separates contemporaneous and lagged climate-index components from the common
trend and seasonal baseline, even when the fitted quantiles themselves are
visually close.

\begin{CodeChunk}
\begin{CodeInput}
R> plot(NA, ylim = c(-1.5, 1.5), xlim = c(2016, 2020), ylab = "covariate contribution")
R> grid(col = "grey90")
R> plot(MREG, type = "component", index = 8:9, add = TRUE, col = "steelblue")
R> plot(MTF, type = "component", index = 8, add = TRUE, col = "forest green")
R> abline(h = 0, col = "orange", lty = 3, lwd = 1.4)
R> legend("topleft", legend = c("MREG direct", "MTF transfer"),
+        col = c("steelblue", "forest green"), lty = 1, lwd = 1.5, bty = "n")
\end{CodeInput}
\end{CodeChunk}

Figure \ref{fig:ex3tftheta} then displays the augmented transfer states
directly. As described in Section \ref{sec:tf}, \(\zeta_t\) controls the
accumulated transfer effect and the \(\psi_t\) states determine how the
contemporaneous covariates enter that transfer term. In this example the
\(\psi_t\) discount factors are set to 1, so the bottom panels summarize static
NOI and AMO coefficients within the transfer-function model. Setting
\code{type = "state"} in \code{plot()} plots a single element of the
augmented state vector \(\tilde{\vect{\theta}}_t\). In this fitted model,
indices 8, 9, and 10 correspond to \(\zeta_t\), \(\psi_{\mathrm{NOI},t}\), and
\(\psi_{\mathrm{AMO},t}\), respectively.

\begin{CodeChunk}
\begin{CodeInput}
R> layout(matrix(c(1, 1, 2, 3), nrow = 2, byrow = TRUE))
R> plot(MTF, type = "state", index = 8, col = "forest green", add = FALSE)
R> grid(col = "grey90")
R> abline(h = 0, col = "orange", lty = 3, lwd = 1.4)
R> title(expression(zeta[t]))
R> par(mar = c(3.8, 4.2, 2.1, 0.8))
R> plot(y.train, type = "n", ylim = c(-0.11, 0.01), xlab = "time", ylab = "component CrIs")
R> plot(MTF, type = "state", index = 9, col = "steelblue", add = TRUE)
R> abline(h = 0, col = "orange", lty = 3, lwd = 1.4)
R> title(expression(psi[list(NOI, t)]))
R> plot(y.train, type = "n", ylim = c(-0.005, 0.06), xlab = "time", ylab = "component CrIs")
R> plot(MTF, type = "state", index = 10, col = "darkorange", add = TRUE)
R> abline(h = 0, col = "orange", lty = 3, lwd = 2)
R> title(expression(psi[list(AMO, t)]))
\end{CodeInput}
\end{CodeChunk}

The fitted object also reports \code{median.kt}, the median number of time steps
until the transfer effect on the \(0.15\) quantile is less than or equal to
\(1\mathrm{e}{-3}\), as discussed in Section \ref{sec:tf}. Here
\code{median.kt} is approximately 25.15, indicating a persistent transfer effect
under the selected \(\lambda\).
\begin{CodeChunk}
\begin{CodeInput}
R> MTF$median.kt
\end{CodeInput}
\begin{CodeOutput}
[1] 25.15187
\end{CodeOutput}
\end{CodeChunk}

We next produce 18-month-ahead forecasts over the held-out window. The
future state-space matrices are passed through the \code{fFF} and \code{fGG}
arguments of the \code{predict()} method for fitted dynamic objects. This
method returns \code{exdqlmForecast} objects.
The
code below shows the relevant matrix construction and forecast calls; the
replication script wraps these steps in dimension checks and output-writing
code. For \code{M0}, the future state-space matrices repeat the trend and
seasonal model.

\begin{CodeChunk}
\begin{CodeInput}
R> k.fore = length(y.holdout)
R> F0.future = model$FF
R> G0.future = model$GG
R> fc.M0 = predict(M0, start.t = length(y.train), k = k.fore, fFF = F0.future,
+                  fGG = G0.future, return.draws = TRUE, n.samp = 1000, seed = 20265101)
\end{CodeInput}
\end{CodeChunk}

For \code{MREG}, the future observation matrix includes the held-out NOI and AMO values as direct regressors.

\begin{CodeChunk}
\begin{CodeInput}
R> n.x = ncol(X.holdout)
R> reg.future = regMod(X.holdout, m0 = rep(0, n.x), C0 = diag(1, n.x))
R> model.reg.future = model + reg.future
R> FREG.future = model.reg.future$FF
R> GREG.future = model.reg.future$GG
R> fc.MREG = predict(MREG, start.t = length(y.train), k = k.fore, fFF = FREG.future,
+                    fGG = GREG.future, return.draws = TRUE, n.samp = 1000, seed = 20265301)
\end{CodeInput}
\end{CodeChunk}

For \code{MTF}, the future matrices include the held-out NOI and AMO values through the transfer-function augmentation.

\begin{CodeChunk}
\begin{CodeInput}
R> n.state = length(model$m0)
R> FTF.future = matrix(0, nrow = n.state + n.x + 1, ncol = k.fore)
R> FTF.future[seq_len(n.state), ] = F0.future
R> FTF.future[n.state + 1, ] = 1
R> GTF.future = array(0, c(n.state + n.x + 1, n.state + n.x + 1, k.fore))
R> GTF.future[seq_len(n.state), seq_len(n.state), ] = G0.future
R> zeta.ind = n.state + 1
R> psi.ind = n.state + 1 + seq_len(n.x)
R> GTF.future[zeta.ind, zeta.ind, ] = lambda.star
R> for (j in seq_len(n.x)) {
+    GTF.future[zeta.ind, psi.ind[j], ] = X.holdout[, j]
+    GTF.future[psi.ind[j], psi.ind[j], ] = 1
+    }
R> fc.MTF = predict(MTF, start.t = length(y.train), k = k.fore, fFF = FTF.future,
+                   fGG = GTF.future, return.draws = TRUE, n.samp = 1000, seed = 20265201)
\end{CodeInput}
\end{CodeChunk}

\begin{figure}[!htb]
\centering
\includegraphics[width=0.82\textwidth]{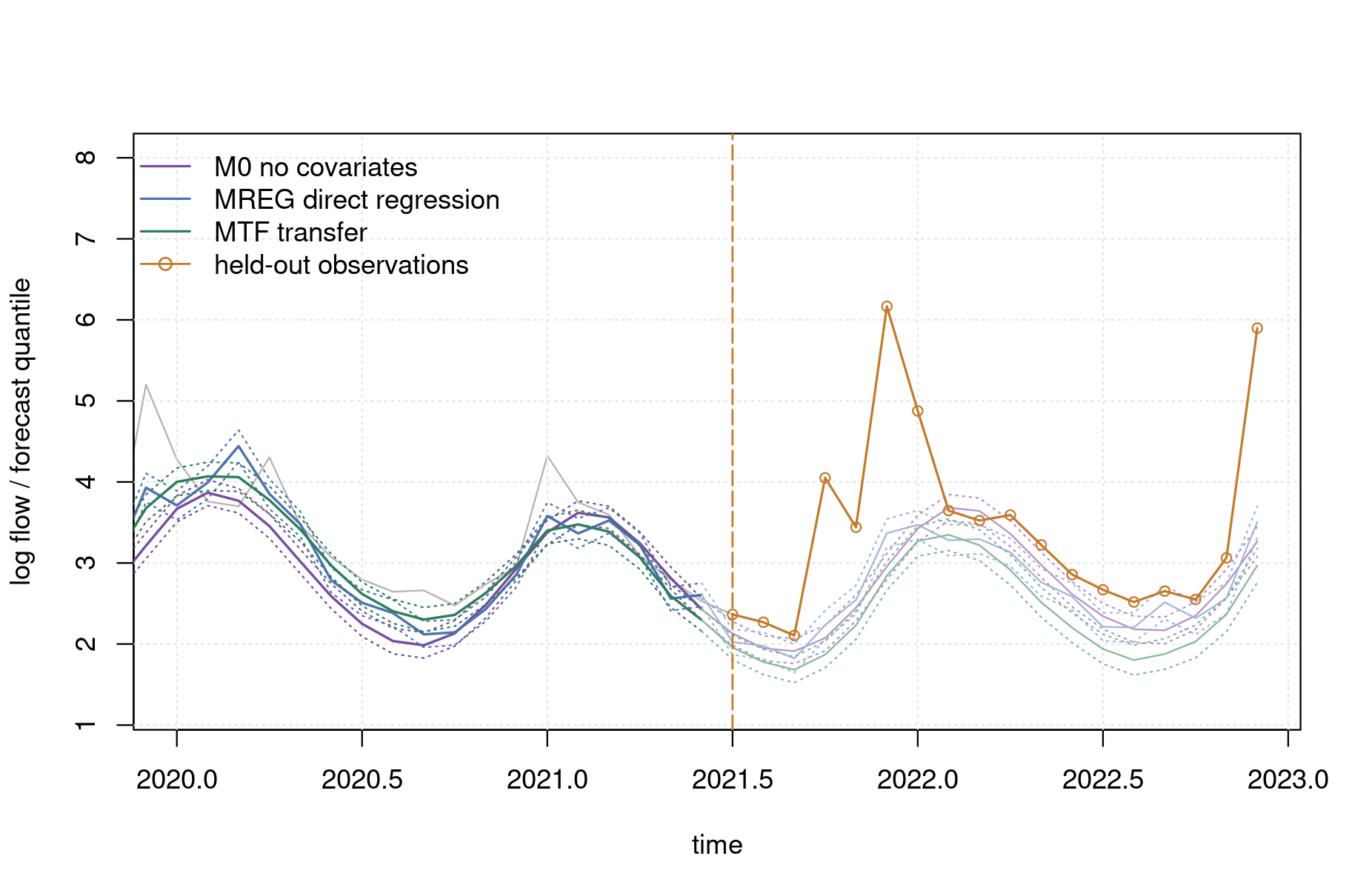}
\caption{Example 3: Big Tree water flow. Held-out 18-month conditional forecast
display from July 2021 through December 2022. The vertical dot-dashed line marks
the forecast origin after the final training month, June 2021. Solid lines
indicate posterior mean forecast quantiles and dotted lines indicate pointwise
95\% CrIs for \code{M0}, \code{MREG}, and \code{MTF}. Held-out observations are shown as orange circles. The forecasts are conditional on the observed holdout-period NOI and AMO values.}
\label{fig:ex3forecast}
\end{figure}

Finally, Figure \ref{fig:ex3forecast} overlays the filtered estimates before the forecast
origin and the conditional forecasts after it.

\begin{CodeChunk}
\begin{CodeInput}
R> plot(y.fit, col = "grey70", xlim = c(2020, 2023), ylim = c(1,8),
        ylab = "log flow / forecast quantile", xlab = "time")
R> grid(col = "grey90")
R> plot(fc.M0, add = TRUE, cols = c("purple", "plum"))
R> plot(fc.MREG, add = TRUE, cols = c("steelblue", "lightblue"))
R> plot(fc.MTF, add = TRUE, cols = c("forest green", "darkseagreen"))
R> t.all = as.numeric(time(y.fit))
R> hold.idx = 415:432
R> lines(t.all[hold.idx], y.fit[hold.idx], col = "darkorange", lwd = 1.4)
R> points(t.all[hold.idx], y.fit[hold.idx], col = "darkorange", pch = 1, cex = 0.8)
R> abline(v = t.all[hold.idx[1]], col = "orange", lty = 5, lwd = 1.2)
R> legend("topleft", legend = c("M0 no covariates", "MREG direct regression",
+            "MTF transfer", "held-out observations"),
+            col = c("purple", "steelblue", "forest green", "darkorange"),
+            lty = 1, pch = c(NA, NA, NA, 1), bty = "n")
\end{CodeInput}
\end{CodeChunk}

The numerical comparison in Table
\ref{tab:ex3} uses \code{diagnostics()} so that all reported quantities
are exported package diagnostics, as defined in Section \ref{sec:diags}. These
diagnostics are computed from the final-training fitted objects.

\begin{CodeChunk}
\begin{CodeInput}
R> diag.M0 = diagnostics(M0)
R> diag.MREG = diagnostics(MREG)
R> diag.MTF = diagnostics(MTF)
R> tab.ex3 = data.frame(model = c("M0", "MREG", "MTF"),
+     KL = c(diag.M0$m1.KL, diag.MREG$m1.KL, diag.MTF$m1.KL),
+     CRPS = c(diag.M0$m1.CRPS, diag.MREG$m1.CRPS, diag.MTF$m1.CRPS),
+     PPLC = c(diag.M0$m1.pplc, diag.MREG$m1.pplc, diag.MTF$m1.pplc))
\end{CodeInput}
\end{CodeChunk}

\begin{table}[ht!]
\centering
\TableStyle
\begin{tabular}{@{}lrrr@{}}
 \toprule
 Model & KL & CRPS & PPLC \\
 \midrule
 \code{M0} no covariates & 0.168 & 0.470 & \textbf{99.427} \\
 \code{MREG} direct regression & \textbf{0.112} & \textbf{0.347} & 151.344 \\
 \code{MTF} transfer function & 0.155 & 0.366 & 125.165 \\
 \bottomrule
\end{tabular}
\caption{Example 3: Big Tree water flow. Package diagnostics from
\code{diagnostics()} for the no-covariate baseline \code{M0}, direct-regression model \code{MREG}, and transfer-function model \code{MTF}, computed on the final-training fitted objects. KL denotes the deterministic Kullback--Leibler normality diagnostic for the MAP standardized forecast errors, CRPS denotes the continuous ranked probability score, and PPLC denotes the posterior predictive loss criterion. Lower values are preferred; bold entries mark the smallest value for each diagnostic.}
\label{tab:ex3}
\end{table}

Alternatively, Table
\ref{tab:ex3forecastmetrics} reports held-out forecast scores over the final 18
months. These diagnostics are computed from the forecast objects using the corresponding \code{diagnostics()} method. The held-out observations are a required input for this method.

\begin{CodeChunk}
\begin{CodeInput}
R> fc.diag.M0 = diagnostics(fc.M0, y = y.holdout)
R> fc.diag.MREG = diagnostics(fc.MREG, y = y.holdout)
R> fc.diag.MTF = diagnostics(fc.MTF, y = y.holdout)
R> tab.ex3.fc = data.frame(model = c("M0", "MREG", "MTF"),
+     check.loss = c(fc.diag.M0$m1.check_loss, fc.diag.MREG$m1.check_loss,
+                    fc.diag.MTF$m1.check_loss),
+     CRPS = c(fc.diag.M0$m1.CRPS, fc.diag.MREG$m1.CRPS, fc.diag.MTF$m1.CRPS))
\end{CodeInput}
\end{CodeChunk}

Table \ref{tab:ex3} gives a mixed in-sample diagnostic comparison. The direct-regression model has the smallest KL and CRPS values, while the no-covariate baseline has the smallest PPLC. The transfer-function model falls between them for PPLC and has diagnostics closer to \code{MREG} than to \code{M0} for CRPS. The example instead shows how \pkg{exdqlm} supports three related dynamic quantile workflows: no external inputs, direct regression through \code{regMod()}, and lagged external-input effects through transfer functions.

\begin{table}[ht!]
\centering
\TableStyle
\begin{tabular}{@{}lrr@{}}
 \toprule
 Model & Check loss & CRPS \\
 \midrule
 \code{M0} no covariates & 0.119 & 0.498 \\
 \code{MREG} direct regression & \textbf{0.114} & \textbf{0.401} \\
 \code{MTF} transfer function & 0.155 & 0.550 \\
 \bottomrule
\end{tabular}
\caption{Example 3: Big Tree water flow. Held-out forecast metrics over July 2021 through December 2022, computed with \code{diagnostics()} from \code{exdqlmForecast} objects returned by \code{predict(..., return.draws = TRUE)}. Check loss is computed at \(p_0=0.15\), and CRPS is computed from posterior predictive forecast draws using the integrated-quantile-score approximation. Lower values are preferred; bold entries mark the smallest value in each column.}
\label{tab:ex3forecastmetrics}
\end{table}

In the held-out comparison, \code{MREG} has the smallest check loss and CRPS, followed by the no-covariate model \code{M0}; \code{MTF} has the largest value for both metrics. Thus, selecting the transfer parameter \(\lambda\) using training-period PPLC does not make the transfer-function specification the best held-out forecasting model in this example. The purpose of \code{MTF} here is to demonstrate how persistent lagged covariate effects can be specified, inspected, and forecast using the package, rather than to claim predictive superiority over direct regression for these data.

\subsection{Static exAL regression on a simulated sparse Gaussian benchmark}
\label{sec:ex4static}

To illustrate the static routines under sparse shrinkage, we consider a correlated Gaussian regression benchmark for which the target conditional quantile is known exactly. Let
\[
\mat{x}_i \sim \N_8(\mat{0}, \Sigma_x),\qquad (\Sigma_x)_{jk} = 0.5^{|j-k|},
\]
with sparse signal
\[
\vect{\beta}^\star = (3,\,1.5,\,0,\,0,\,2,\,0,\,0,\,0)^\top.
\]
We use the following benchmark data-generating mechanism.
For each target quantile level \(p_0\), we generate
\[
Y_i^{(p_0)} = \mat{x}_i^\top \vect{\beta}^\star + \sigma_\varepsilon \{Z_i - \Phi^{-1}(p_0)\},
\qquad Z_i \sim \N(0,1),\qquad \sigma_\varepsilon = 1.5.
\]
Then
\[
Q_Y(p_0 \mid \mat{x}_i) = \mat{x}_i^\top \vect{\beta}^\star,
\]
so the true \(p_0\)-quantile is the same sparse linear signal at each fitted quantile level. Note that this is a quantile-specific calibration benchmark: each target level is generated so that the known target quantile is the same sparse linear signal, rather than a single-response multi-quantile analysis. The Gaussian data-generating distribution does not coincide with the exAL working likelihood; the benchmark therefore evaluates recovery of a correctly specified conditional quantile under likelihood misspecification. Gaussian design matrices are generated with \code{MASS::mvrnorm()} from \pkg{MASS} \citep{MASS}.
We use \(n = 160\) training observations and an independent holdout sample of size \(800\).

\begin{CodeChunk}
\begin{CodeInput}
R> Sigma.x = 0.5 ^ as.matrix(dist(1:8))
R> beta.true = c(3, 1.5, 0, 0, 2, 0, 0, 0)
R> rhs.ctrl = list(tau0 = 0.15, zeta2_fixed = 9,
+                  shrink_intercept = FALSE)
R> set.seed(20260712)
R> sim.y = function(X.raw, p0) {
+     drop(X.raw %*% beta.true) + 1.5 * (rnorm(nrow(X.raw)) - qnorm(p0))}
R> X.raw = MASS::mvrnorm(160, mu = rep(0, 8), Sigma = Sigma.x)
R> X = cbind(1, X.raw)
R> X.hold.raw = MASS::mvrnorm(800, mu = rep(0, 8), Sigma = Sigma.x)
R> X.hold = cbind(1, X.hold.raw)
R> ref.hold = drop(X.hold %*% c(0, beta.true))
R> p.grid = c(0.05, 0.25, 0.50)
\end{CodeInput}
\end{CodeChunk}

We fit separate static exAL models at \(p_0 \in \{0.05,0.25,0.50\}\). All fits use the Nishimura--Suchard regularized horseshoe prior \citep{nishimura2023shrunken}, implemented in the package through \code{beta_prior = "rhs\_ns"}, with \(\tau_0 = 0.15\), fixed slab \(\zeta^2 = 9\), and an unshrunk intercept. We fit an LDVB approximation with \code{exalStaticLDVB()} using \code{n.samp = 3000}, and an MCMC fit with \code{exalStaticMCMC()} using \code{n.burn = 2000} and \code{n.mcmc = 3000}; the MCMC routine uses its default \code{slice} kernel and is warm-started from the LDVB fit. The lower-tail LDVB fit at \(p_0 = 0.05\) converges more slowly, so we allow a larger LDVB iteration budget there. Setting \code{al.ind = TRUE} (or equivalently \code{dqlm.ind = TRUE}) would recover the AL special case discussed in Section \ref{sec:exdqlm}, but here we focus on the general static exAL model under sparse \code{rhs\_ns} shrinkage.

\begin{CodeChunk}
\begin{CodeInput}
R> y.train = y.hold = M.ldvb = M.mcmc = vector("list", length(p.grid))
R> for (i in seq_along(p.grid)) {
+     p0 = p.grid[i]
+     y.train[[i]] = sim.y(X.raw, p0)
+     y.hold[[i]] = sim.y(X.hold.raw, p0)
+     M.ldvb[[i]] = exalStaticLDVB(y = y.train[[i]], X = X, p0 = p0,
+                    beta_prior = "rhs_ns", beta_prior_controls = rhs.ctrl,
+                    max_iter = ifelse(p0 == 0.05, 420, 260),
+                    n.samp = 3000, tol = 1e-4, verbose = FALSE)
+     M.mcmc[[i]] = exalStaticMCMC(y = y.train[[i]], X = X, p0 = p0,
+                    beta_prior = "rhs_ns", beta_prior_controls = rhs.ctrl,
+                    n.burn = 2000, n.mcmc = 3000, thin = 1,
+                    init.from.vb = TRUE, verbose = FALSE)}
\end{CodeInput}
\end{CodeChunk}

\begin{figure}[!t]
\centering
\includegraphics[width=0.95\textwidth]{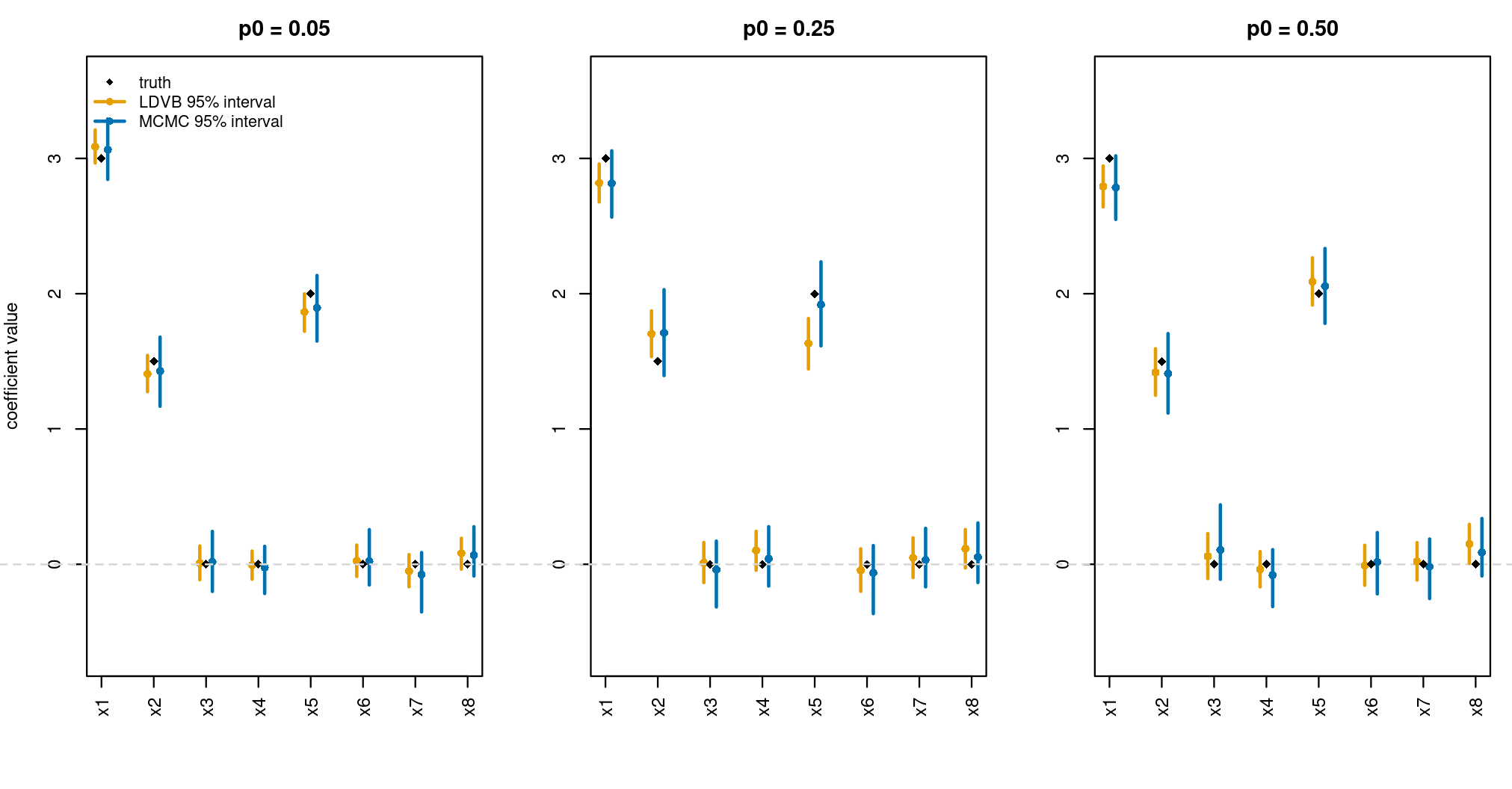}
\caption{Example 4: sparse static exAL regression under the correlated Gaussian benchmark with the \code{rhs\_ns} prior. Black diamonds denote the true non-intercept coefficients, orange intervals denote LDVB 95\% posterior intervals, and blue intervals denote MCMC 95\% posterior intervals. Panels correspond to \(p_0 = 0.05, 0.25,\) and \(0.50\).}
\label{fig:ex4static}
\end{figure}

For static fits, the \code{diagnostics()} method summarizes fitted quantiles on a common design matrix and, when holdout responses or a reference quantile are supplied, reports predictive evaluation metrics such as check loss and reference-quantile error. It also stores posterior coefficient summaries that can be plotted with \code{plot(..., type = "coefficients")}. In this simulated example, the true coefficient vector is known, so Figure \ref{fig:ex4static} overlays it as a visual reference. This truth overlay is optional through the parameter \code{beta.ref} and is not needed for real data applications.

\begin{CodeChunk}
\begin{CodeInput}
R> diag.static = Map(function(ldvb, mcmc, y.h) {
+     diagnostics(ldvb, mcmc, X = X.hold, y = y.h, ref = ref.hold)
+   }, M.ldvb, M.mcmc, y.hold)
R> y.lim = range(beta.true,
+                unlist(lapply(diag.static, function(z){
+                  c(z$m1.beta.lb[-1], z$m1.beta.ub[-1], z$m2.beta.lb[-1],
+                  z$m2.beta.ub[-1])})))
R> y.lim = y.lim + c(-1, 1) * 0.08 * diff(y.lim)
R> par(mfrow = c(1, 3), mar = c(5.2, 4.0, 2.6, 1.0), xpd = NA)
R> for (i in seq_along(p.grid)) {plot(diag.static[[i]], type = "coefficients",
+          beta.ref = c(0, beta.true), include.intercept = FALSE,
+          coef.names = c("(Intercept)", paste0("x", seq_along(beta.true))),
+          cols = c("orange", "steelblue"), legend.labels = c("LDVB 95% interval",
+          "MCMC 95% interval"), beta.ref.label = "truth", ylim = y.lim,
+          ylab = if (i == 1) "coefficient value" else "",
+          main = sprintf("p0 = %.2f", p.grid[i]), legend = i == 1)}
\end{CodeInput}
\end{CodeChunk}

Figure \ref{fig:ex4static} shows that, in this simulated realization, both methods recover the main sparse signal structure and shrink most null coefficients toward zero across the lower tail, lower-middle quantile, and median. Recovery is not uniform across coefficients: the smaller active coefficient is more difficult in the median fit, and a few null-coefficient intervals can miss zero. Across the three panels, the MCMC intervals tend to give slightly cleaner active-signal recovery, while the LDVB intervals provide a fast approximation to the same coefficient-level summaries.

\begin{table}[t!]
\centering
\TableStyle
{\setlength{\tabcolsep}{3pt}
\begin{tabular}{@{}llrrrr@{}}
 \toprule
 \(p_0\) & method & \shortstack{runtime\\(s)} & \shortstack{active\\RMSE} & \shortstack{null\\MAE} & \shortstack{holdout\\qRMSE} \\
 \midrule
 \tworowcell{0.05} & LDVB & \textbf{29.81} & 0.107 & \textbf{0.034} & 0.674 \\
 & MCMC & 211.27 & \textbf{0.082} & 0.041 & \textbf{0.586} \\
 \midrule
 \tworowcell{0.25} & LDVB & \textbf{17.95} & 0.264 & 0.064 & 0.410 \\
 & MCMC & 216.40 & \textbf{0.168} & \textbf{0.046} & \textbf{0.280} \\
 \midrule
\tworowcell{0.50} & LDVB & \textbf{10.57} & \textbf{0.138} & \textbf{0.055} & \textbf{0.306} \\
 & MCMC & 207.98 & 0.139 & 0.062 & 0.331 \\
 \bottomrule
\end{tabular}}
\caption{Example 4: runtime and sparse-signal recovery metrics for the static exAL fits under the \code{rhs\_ns} prior. The LDVB rows use \code{n.samp = 3000}, and the MCMC rows use 2000 burn-in iterations and 3000 retained draws. The active RMSE is computed over the three nonzero coefficients, the null MAE is computed over the five zero coefficients, and the holdout quantile RMSE is computed against the true target quantile on an independent holdout sample of size 800. Because the columns are the comparison metrics, bold entries indicate the smaller value within each quantile block; displayed ties are left unbolded.}
\label{tab:ex4static}
\end{table}

Table \ref{tab:ex4static} includes active-signal RMSE and null-coefficient MAE summaries stored in the \code{exalStaticDiagnostic} objects returned by \code{diagnostics()}. The results reinforce the visual comparison in Figure \ref{fig:ex4static}. The MCMC posterior-simulation fit has the smaller holdout quantile RMSE at \(p_0 = 0.05\) and \(p_0 = 0.25\), while LDVB is smaller at \(p_0 = 0.50\). Active-signal recovery shows the same pattern for \(p_0 = 0.05\) and \(p_0 = 0.25\), with LDVB slightly smaller at \(p_0 = 0.50\). The LDVB fit is nevertheless substantially faster in all three cases, with runtime reductions of roughly 7-fold at \(p_0 = 0.05\), 12-fold at \(p_0 = 0.25\), and 20-fold at \(p_0 = 0.50\). In this benchmark, the LDVB approximation is useful as a rapid first-pass screen, while static MCMC provides a more expensive posterior-simulation comparison when interval accuracy or posterior tail behavior is central.

\section{Conclusion}
\label{sec:summary}

\pkg{exdqlm} provides an \proglang{R} workflow for Bayesian quantile regression with primary emphasis on latent state-space models fitted one response quantile at a time. It addresses a gap between broad quantile-regression tools, which do not provide the dedicated latent quantile-state workflow targeted here, and Gaussian dynamic-modeling packages, which do not target AL or exAL quantile likelihoods. Within this setting, \pkg{exdqlm} fits DQLMs and exDQLMs, offers MCMC posterior simulation and LDVB approximate inference, and uses the resulting fitted objects for state and component summaries, forecasting, calibration and predictive diagnostics including deterministic KL, CRPS, and PPLC, and posterior-predictive synthesis across separately fitted quantile levels. The package also includes transfer-function state augmentation and static AL/exAL regression with shrinkage priors. These capabilities are exposed through composable model constructors and fitted objects supporting standard methods such as \code{summary()}, \code{plot()}, \code{predict()}, and \code{diagnostics()}.

The examples illustrate how these interfaces are used in analysis. Lake Huron uses separate dynamic fits at three quantile levels, multi-step forecasting, and post hoc synthesis of posterior predictive draws. Sunspots compares DQLM and exDQLM fits at an upper quantile, uses MAP plug-in one-step-ahead calibration diagnostics together with CRPS and PPLC, and illustrates an in-sample comparison of candidate discount-factor specifications using CRPS, with KL used as a complementary calibration check. Big Tree water flow compares no-covariate, direct-regression, and transfer-function specifications built from a common trend-seasonal baseline; its fitted and held-out summaries are mixed, so the example is mainly a workflow comparison rather than evidence for a uniformly preferred covariate structure. The static example uses one quantile-specific sparse simulation realization to illustrate LDVB and MCMC under a known target quantile.

The examples also delineate the current scope of the package. Dynamic models are fitted one quantile level at a time. \code{quantileSynthesis()} combines predictive draws after separate fits and applies monotonicity adjustments when needed; it is not a joint posterior model over multiple quantile levels. LDVB returns samples from fitted variational factors rather than from the exact posterior target, so MCMC remains the reference calculation when posterior tail summaries, interval widths, or approximation error are central. In transfer-function models, lag propagation is controlled by a fixed or grid-selected \(\lambda\); the current implementation does not estimate input-specific lag-decay rates internally. Runtime results are elapsed fitting times under the stated backend profile and should be read alongside predictive and calibration diagnostics.

Several development directions follow from these limitations: multivariate-output dynamic quantile models that represent cross-series dependence through a modeled correlation matrix for related time series; improved approximations for the nonconjugate joint \((\gamma,\sigma)\) skewness--scale block; nonlinear dynamic quantile modeling beyond the linear state-space form used here; and joint or regularized fitting across quantile levels. The main text, appendices, and replication materials have separate roles: the main text describes the user-facing workflow, the appendices record selected posterior targets and package-specific computational blocks, and the repository scripts regenerate the reported figures, tables, example artifacts, and benchmark provenance.

\clearpage
\appendix
\exdqlmAppendixSetup

\section{Technical overview and notation}
\label{sec:app_scope}

The appendix aims to connect the package interface discussed in the article with the
posterior targets and computations implemented in \pkg{exdqlm}. It records the
distributional conventions, compact dynamic and static posterior targets, the
Laplace--delta variational Bayes (LDVB) scale--skewness block, the static
\code{rhs\_ns} shrinkage block, diagnostic and scoring quantities, and the
posterior-predictive synthesis algorithm. Standard Gaussian filtering,
smoothing, and backward-sampling recursions are cited rather than rederived.
Function-level arguments and backend controls are documented in the package
help pages, and the replication materials provide executable code for the
reported figures, tables, and benchmark quantities.

\subsection{Distributional conventions}
\label{sec:app_distributional_conventions}

Let $p_0\in(0,1)$ be the target quantile. The normal distribution
$\N(m,V)$ is parameterized by its mean and variance. The inverse-gamma
distribution uses the shape-rate convention
\begin{equation}
x\sim\IG(a,b)
\quad\Longleftrightarrow\quad
p(x)=\frac{b^a}{\Gamma(a)}x^{-a-1}\exp(-b/x),\qquad x>0.
\end{equation}
The exponential distribution $\Exp(\text{rate}=1/\sigma)$ has mean $\sigma$.
The notation $\TNp(m,V)$ denotes $\N(m,V)$ truncated to $(0,\infty)$.

The asymmetric Laplace (AL) mixture representation used by the package is
\begin{align}
v &\mid \sigma \sim \Exp(\text{rate}=1/\sigma),\\
y \mid \eta,\sigma,v
&\sim \N\{ \eta + A(p_0)v,\ \sigma B(p_0)v \},
\end{align}
where $\eta$ is the target-quantile location and
\begin{equation}
A(p)=\frac{1-2p}{p(1-p)}, \qquad B(p)=\frac{2}{p(1-p)}.
\label{eq:al_AB}
\end{equation}

The extended asymmetric Laplace (exAL) representation adds a latent
$s\sim\TNp(0,1)$. For $\gamma\in(L,U)$, define
\begin{align}
g(\gamma) &= 2\Phi(-|\gamma|)\exp(\gamma^2/2),\\
p_\gamma &= \Ind(\gamma<0) +
  \frac{p_0-\Ind(\gamma<0)}{g(\gamma)},\\
A_\gamma &= A(p_\gamma),\qquad
B_\gamma = B(p_\gamma),\qquad
C_\gamma = \{\Ind(\gamma>0)-p_\gamma\}^{-1}.
\label{eq:exal_ABC}
\end{align}
The bounds $L<0<U$ are the roots $g(L)=1-p_0$ and $g(U)=p_0$. Conditional on
$(v,s,\sigma,\gamma)$,
\begin{equation}
y\mid \eta,\sigma,\gamma,v,s
\sim
\N\left(\eta + C_\gamma\sigma|\gamma|s + A_\gamma v,\
\sigma B_\gamma v \right).
\label{eq:exal_aug}
\end{equation}
The package uses $\sigma\sim\IG(a_\sigma,b_\sigma)$ and a proper truncated
Student-$t$ prior $\pigamma(\gamma)$ for $\gamma$ on $(L,U)$:
\begin{equation}
\pigamma(\gamma)
\propto
s_\gamma^{-1}t_{\nu_\gamma}\{(\gamma-m_\gamma)/s_\gamma\}
\Ind(L<\gamma<U),
\label{eq:gamma_prior}
\end{equation}
with the normalizing constant determined by the Student-$t$ distribution
function over $(L,U)$.

The generalized inverse Gaussian (GIG) distribution is parameterized as
\begin{equation}
x\sim\GIG(\lambda,\chi,\psi_{\mathrm{GIG}})
\quad\Longleftrightarrow\quad
p(x)\propto x^{\lambda-1}
\exp\left\{-\frac{1}{2}\left(\frac{\chi}{x}
+\psi_{\mathrm{GIG}} x\right)\right\},\qquad x>0.
\label{eq:gig_parameterization}
\end{equation}
For $r\in\R$,
\begin{equation}
\E(x^r)=
\left(\frac{\chi}{\psi_{\mathrm{GIG}}}\right)^{r/2}
\frac{K_{\lambda+r}(\sqrt{\chi\psi_{\mathrm{GIG}}})}
{K_{\lambda}(\sqrt{\chi\psi_{\mathrm{GIG}}})},
\label{eq:gig_moments}
\end{equation}
where $K_\nu(\cdot)$ is the modified Bessel function of the second kind.

\section{Posterior targets and model blocks}

\subsection{Dynamic DQLM under the AL likelihood}
\label{sec:dyn_dqlm}

For the dynamic quantile linear model (DQLM), the observation and evolution
equations are
\begin{align}
y_t\mid\vect{\theta}_t,\sigma,v_t
&\sim
\N\left(\mat{F}_t^\top\vect{\theta}_t+A(p_0)v_t,\
\sigma B(p_0)v_t\right),\\
\vect{\theta}_t\mid\vect{\theta}_{t-1}
&\sim \N(\mat{G}_t\vect{\theta}_{t-1},\mat{W}_t),\\
v_t\mid\sigma &\sim \Exp(\text{rate}=1/\sigma).
\label{eq:dyn_dqlm_hierarchy}
\end{align}
The posterior target is proportional to
\begin{equation}
p(\vect{\theta}_{0:T},\vect{v},\sigma\mid\vect{y})
\propto
p(\vect{\theta}_0)p(\sigma)
\prod_{t=1}^T
p(y_t\mid\vect{\theta}_t,\sigma,v_t)
p(v_t\mid\sigma)
p(\vect{\theta}_t\mid\vect{\theta}_{t-1}).
\label{eq:dyn_dqlm_joint}
\end{equation}
Conditioning on $(\sigma,\vect{v})$, the state sequence is a Gaussian
state-space model with observation offset $A(p_0)v_t$ and variance
$\sigma B(p_0)v_t$. The package uses standard Kalman filtering,
Rauch--Tung--Striebel smoothing, and forward-filtering backward-sampling
recursions for this Gaussian block \citep{west1997bayesian}.
\label{sec:gaussian_state_update}

\subsection{Dynamic exDQLM under the exAL likelihood}
\label{sec:dyn_exdqlm}

The dynamic exDQLM augments the AL hierarchy with $s_t\sim\TNp(0,1)$ and
the exAL constants in \eqref{eq:exal_ABC}:
\begin{align}
y_t\mid\vect{\theta}_t,\sigma,\gamma,v_t,s_t
&\sim
\N\left(
\mat{F}_t^\top\vect{\theta}_t
+C_\gamma\sigma|\gamma|s_t+A_\gamma v_t,\
\sigma B_\gamma v_t
\right),\\
\vect{\theta}_t\mid\vect{\theta}_{t-1}
&\sim \N(\mat{G}_t\vect{\theta}_{t-1},\mat{W}_t),\\
v_t\mid\sigma&\sim\Exp(\text{rate}=1/\sigma),\qquad
s_t\sim\TNp(0,1).
\label{eq:dyn_exdqlm_obs}
\end{align}
The posterior target is
\begin{align}
p(\vect{\theta}_{0:T},\vect{v},\vect{s},\sigma,\gamma\mid\vect{y})
&\propto
p(\vect{\theta}_0)p(\sigma)\pigamma(\gamma)
\prod_{t=1}^T p(y_t\mid\vect{\theta}_t,\sigma,\gamma,v_t,s_t)
\nonumber\\
&\quad{}\times
\prod_{t=1}^T
p(v_t\mid\sigma)p(s_t)
p(\vect{\theta}_t\mid\vect{\theta}_{t-1}).
\label{eq:dyn_exdqlm_joint}
\end{align}
Given $(\sigma,\gamma,\vect{v},\vect{s})$, the state sequence is again a
Gaussian state-space model, now with offset
$A_\gamma v_t+C_\gamma\sigma|\gamma|s_t$ and variance
$\sigma B_\gamma v_t$.

\subsection{Transfer-function state augmentation}
\label{sec:appendix_transfer_function}

Appendix notation follows the transfer-function formulation in Section
\ref{sec:tf}. Let \(\mat{x}_t=(x_{1t},\ldots,x_{kt})^\top\) denote the
\(k\)-dimensional input vector. The package augments the base dynamic state with
a scalar accumulated transfer effect \(\zeta_t\) and a vector of instantaneous
transfer coefficients
\[
\vect{\psi}_t=(\psi_{1,t},\ldots,\psi_{k,t})^\top.
\]
The augmented state is
\begin{equation}
\vect{\theta}^{\mathrm{aug}}_t
=
(\vect{\theta}^{\mathrm{base}\top}_t,\zeta_t,\vect{\psi}_t^\top)^\top,
\label{eq:tf_aug_state}
\end{equation}
with observation row
\[
\tilde{\mat{F}}_t^\top=(\mat{F}_t^\top,1,\mat{0}_k^\top).
\]
For fixed transfer rate \(\lambda\), the transfer block of the evolution matrix is
\[
\mat{G}^{tf}_t =
\begin{pmatrix}
\lambda & \mat{x}_t^\top \\
\mat{0}_k & \mat{I}_k
\end{pmatrix},
\]
so that the scalar accumulated effect evolves as
\(\zeta_t=\lambda\zeta_{t-1}+\mat{x}_t^\top\vect{\psi}_{t-1}+\omega_t\),
while \(\vect{\psi}_t\) follows a random walk with evolution variance specified
through the transfer-function discount factors. In Example \ref{sec:ex3},
\(k=2\), with normalized NOI and AMO as transfer inputs; the appended transfer
states are therefore \(\zeta_t\), \(\psi_{\mathrm{NOI},t}\), and
\(\psi_{\mathrm{AMO},t}\).

After this state augmentation, the fitting algorithms are the same dynamic MCMC
or LDVB routines used for the corresponding exDQLM.

\subsection{Static AL and exAL regression}
\label{sec:static_ridge}

Let $\mat{X}\in\R^{n\times p}$ have rows $\vect{x}_i^\top$ and regression
coefficient vector $\vect{\beta}$. Under a Gaussian/ridge prior,
\begin{equation}
\vect{\beta}\sim\N(\vect{b}_{\beta,0},\mat{V}_{\beta,0}).
\label{eq:static_beta_ridge}
\end{equation}
The static exAL likelihood is
\begin{align}
y_i\mid\vect{\beta},\sigma,\gamma,v_i,s_i
&\sim
\N\left(
\vect{x}_i^\top\vect{\beta}
+C_\gamma\sigma|\gamma|s_i+A_\gamma v_i,\
\sigma B_\gamma v_i
\right),\\
v_i\mid\sigma&\sim \Exp(\text{rate}=1/\sigma),\qquad
s_i\sim\TNp(0,1).
\label{eq:static_exal_hierarchy}
\end{align}
The posterior target is
\begin{align}
p(\vect{\beta},\sigma,\gamma,\vect{v},\vect{s}\mid\vect{y})
&\propto
p(\vect{\beta})p(\sigma)\pigamma(\gamma)
\prod_{i=1}^n
p(y_i\mid\vect{\beta},\sigma,\gamma,v_i,s_i)
p(v_i\mid\sigma)p(s_i).
\label{eq:static_exal_target}
\end{align}
The AL static regression is recovered by fixing $\gamma=0$ and dropping the
$s_i$ block. For fixed latent variables and scale parameters, the
coefficient update is the weighted Gaussian regression update; for
\code{beta_prior = "rhs\_ns"}, the Gaussian prior precision is replaced by
the expected \code{rhs\_ns} precision described in
Section~\ref{sec:static_rhs}.

\section{Laplace--delta VB implementation for exAL models}
\label{sec:ldvb_details}

The nonconjugate part of the exAL posterior is the joint scale--skewness
block $(\sigma,\gamma)$. The LDVB routines in \pkg{exdqlm} specialize the
Laplace--delta strategy of \citet{wang2013nonconjugatevb} to this block.
The dynamic variational family is
\begin{equation}
q(\vect{\theta}_{0:T},\vect{v},\vect{s},\sigma,\gamma)
=q(\vect{\theta}_{0:T})\prod_{t=1}^Tq(v_t)\prod_{t=1}^Tq(s_t)\,
q(\sigma,\gamma).
\label{eq:dyn_exdqlm_mf}
\end{equation}
The static family replaces $q(\vect{\theta}_{0:T})$ by
$q(\vect{\beta})$:
\begin{equation}
q(\vect{\beta},\vect{v},\vect{s},\sigma,\gamma)
=q(\vect{\beta})\prod_{i=1}^nq(v_i)\prod_{i=1}^nq(s_i)\,
q(\sigma,\gamma).
\label{eq:static_exal_mf}
\end{equation}

The conjugate variational factors require moments of functions of
$(\sigma,\gamma)$. In the dynamic case, define
\begin{align}
\kappa_1 &= \E\{1/(\sigma B_\gamma)\},&
\kappa_2 &= \E\{A_\gamma/(\sigma B_\gamma)\},&
\kappa_3 &= \E\{A_\gamma^2/(\sigma B_\gamma)\},\\
\kappa_4 &= \E\{C_\gamma|\gamma|/B_\gamma\},&
\kappa_5 &= \E\{\sigma C_\gamma^2\gamma^2/B_\gamma\},&
\kappa_6 &= \E\{A_\gamma C_\gamma|\gamma|/B_\gamma\}.
\label{eq:dyn_exdqlm_kappa}
\end{align}
For the dynamic state factor, let $m_{1/v,t}=\E(v_t^{-1})$ and
$\bar s_t=\E(s_t)$. The Gaussian pseudo-observation has precision
\begin{equation}
\phi_t=\kappa_1m_{1/v,t}
\label{eq:dyn_exdqlm_vb_precision}
\end{equation}
and pseudo-response
\begin{equation}
y_t^{\mathrm{VB}}
=
\frac{
y_t\phi_t-\kappa_2-\kappa_4\bar s_t m_{1/v,t}
}{\phi_t}.
\label{eq:dyn_exdqlm_vb_pseudo}
\end{equation}
Thus the state factor is updated by the same Gaussian state-space recursions
as in Section~\ref{sec:dyn_dqlm}.

For static regression, the coefficient factor is Gaussian. With row vectors
$\vect{x}_i^\top$,
\begin{align}
\omega_i&=\kappa_1\E(v_i^{-1}),\\
\eta_i&=
y_i\kappa_1\E(v_i^{-1})
-\kappa_4\E(v_i^{-1})\E(s_i)
-\kappa_2,\\
\mat{\Sigma}_\beta
&=\left(\mat{V}_{\beta,0}^{-1}
+\mat{X}^\top\diag(\omega_1,\ldots,\omega_n)\mat{X}\right)^{-1},\\
\vect{m}_\beta
&=\mat{\Sigma}_\beta
\left(\mat{V}_{\beta,0}^{-1}\vect{b}_{\beta,0}
+\mat{X}^\top\vect{\eta}\right),
\label{eq:static_exal_vb_beta}
\end{align}
where $\vect{\eta}=(\eta_1,\ldots,\eta_n)^\top$.

The latent-variable factors remain closed form conditional on the
Laplace--delta moments. If $m_{\eta,t}=\E(\mat{F}_t^\top\vect{\theta}_t)$,
$S_{\eta,t}=\Var(\mat{F}_t^\top\vect{\theta}_t)$, and
$S_{\Delta,t}=(y_t-m_{\eta,t})^2+S_{\eta,t}$, then the dynamic updates are
\begin{align}
q(v_t)&=\GIG\left(
\frac{1}{2},\
\kappa_1S_{\Delta,t}-2\kappa_4(y_t-m_{\eta,t})\E(s_t)
+\kappa_5\E(s_t^2),\
\kappa_3+2\E(\sigma^{-1})
\right),\\
q(s_t)&=\TNp(\mu_{s,t}^{\mathrm{VB}},V_{s,t}^{\mathrm{VB}}),\\
V_{s,t}^{\mathrm{VB}}
&=\left(1+\kappa_5\E(v_t^{-1})\right)^{-1},\\
\mu_{s,t}^{\mathrm{VB}}
&=V_{s,t}^{\mathrm{VB}}
\left\{\kappa_4(y_t-m_{\eta,t})\E(v_t^{-1})-\kappa_6\right\}.
\label{eq:dyn_exdqlm_vb_vs}
\end{align}
The static updates replace $m_{\eta,t}$ and $S_{\eta,t}$ by
$\vect{x}_i^\top\vect{m}_\beta$ and
$\vect{x}_i^\top\mat{\Sigma}_\beta\vect{x}_i$.

The nonconjugate factor is approximated on transformed coordinates
\begin{equation}
\ell_\sigma=\log\sigma,\qquad
z_\gamma=\logit\left(\frac{\gamma-L}{U-L}\right).
\label{eq:dyn_exdqlm_ld_transform}
\end{equation}
If $r=\expit(z_\gamma)$, the transformed objective is
\begin{align}
h_z(\ell_\sigma,z_\gamma)
&=
\E_{-(\sigma,\gamma)}
\{\log p(\vect{y},\text{latent variables},\text{parameters})\}
\nonumber\\
&\quad
+\ell_\sigma+\log(U-L)+\log r+\log(1-r).
\label{eq:dyn_exdqlm_ld_objective}
\end{align}
The package maximizes this objective, uses the local Hessian to form a
Gaussian approximation on $(\ell_\sigma,z_\gamma)$, and evaluates needed
expectations by the delta method. The returned LDVB objects include the
transformed mode, approximate covariance, derived moments, ELBO trace, and
approximate posterior draws when requested.

The dynamic and static exAL ELBOs have the same block structure:
\begin{equation}
\elbo(q)=
\elbo_{\mathrm{like}}+\elbo_{\mathrm{Gaussian}}
+\elbo_v+\elbo_s+\elbo_{\sigma\gamma}
+H(q_{\mathrm{Gaussian}})
+\sum H\{q(v)\}+\sum H\{q(s)\}+H\{q(\sigma,\gamma)\}.
\label{eq:dyn_exdqlm_elbo}
\end{equation}
For dynamic models, the Gaussian block is the state trajectory
$q(\vect{\theta}_{0:T})$; for static models, it is $q(\vect{\beta})$.
The entropy terms use the distributional conventions stated in
Section~\ref{sec:app_distributional_conventions}. For AL/DQLM fits,
$\gamma=0$, the $s$ block is omitted, and $q(\sigma)$ is inverse-gamma.

\section{Static regression with the Nishimura--Suchard regularized horseshoe}
\label{sec:static_rhs}

The \code{rhs\_ns} option implements the conditionally conjugate
regularized horseshoe formulation of \citet{nishimura2023shrunken} for
static regression. Let $\mathcal{A}\subseteq\{1,\ldots,p\}$ be the set of
coefficients subject to shrinkage; by default an intercept is not shrunk. For
$j\in\mathcal{A}$,
\begin{align}
\beta_j\mid\tau^2,\lambda_j^2,\zeta^2
&\propto
\N(\beta_j\mid0,\tau^2\lambda_j^2)
\N(\beta_j\mid0,\zeta^2),\\
\lambda_j^2\mid\nu_j &\sim \IG(1/2,1/\nu_j),&
\nu_j&\sim\IG(1/2,1),\\
\tau^2\mid\xi &\sim\IG(1/2,1/\xi),&
\xi&\sim\IG(1/2,1/\tau_0^2),\\
\zeta^2&\sim\IG(a_\zeta,b_\zeta),
\label{eq:rhs_ns_hierarchy}
\end{align}
unless the slab variance $\zeta^2$ is fixed by the user. The active prior
precision contribution is
\begin{equation}
\omega^{\mathrm{prior}}_j
=\frac{1}{\tau^2\lambda_j^2}+\frac{1}{\zeta^2}.
\label{eq:rhs_ns_precision}
\end{equation}
The static Gaussian coefficient update uses this precision, or its
variational expectation, on the active diagonal entries.

Conditioning on $\vect{\beta}$, the shrinkage MCMC block has closed-form
updates:
\begin{align}
\lambda_j^2\mid\cdot
&\sim
\IG\left(1,\ \frac{1}{\nu_j}+\frac{\beta_j^2}{2\tau^2}\right),\\
\nu_j\mid\cdot
&\sim
\IG\left(1,\ 1+\frac{1}{\lambda_j^2}\right),\\
\tau^2\mid\cdot
&\sim
\IG\left(\frac{|\mathcal{A}|+1}{2},\
\frac{1}{\xi}
+\frac{1}{2}\sum_{j\in\mathcal{A}}\frac{\beta_j^2}{\lambda_j^2}
\right),\\
\xi\mid\cdot
&\sim
\IG\left(1,\ \frac{1}{\tau_0^2}+\frac{1}{\tau^2}\right),\\
\zeta^2\mid\cdot
&\sim
\IG\left(a_\zeta+\frac{|\mathcal{A}|}{2},\
b_\zeta+\frac{1}{2}\sum_{j\in\mathcal{A}}\beta_j^2
\right),
\label{eq:rhs_ns_mcmc}
\end{align}
with the final line omitted when $\zeta^2$ is fixed.

The mean-field VB block has inverse-gamma factors
\[
q(\lambda_j^2),\quad q(\nu_j),\quad q(\tau^2),\quad q(\xi),
\quad q(\zeta^2)\ \text{if }\zeta^2\text{ is random}.
\]
Let $\bar\beta_j^2=\E_q(\beta_j^2)$. The rate updates are
\begin{align}
b_{\lambda j}
&=\frac{1}{2}\bar\beta_j^2\,\E\{(\tau^2)^{-1}\}
+\E(\nu_j^{-1}),\\
b_{\nu j}
&=1+\E\{(\lambda_j^2)^{-1}\},\\
b_\tau
&=\frac{1}{2}\sum_{j\in\mathcal{A}}\bar\beta_j^2
\E\{(\lambda_j^2)^{-1}\}+\E(\xi^{-1}),\\
b_\xi
&=\frac{1}{\tau_0^2}+\E\{(\tau^2)^{-1}\},\\
a_\zeta^\star
&=a_\zeta+\frac{|\mathcal{A}|}{2},\qquad
b_\zeta^\star=b_\zeta+\frac{1}{2}\sum_{j\in\mathcal{A}}\bar\beta_j^2.
\label{eq:rhs_ns_vb_updates}
\end{align}
The shape parameters are
$a_{\lambda j}=a_{\nu j}=a_\xi=1$ and
$a_\tau=(|\mathcal{A}|+1)/2$.

The \code{rhs\_ns} ELBO is the sum of coefficient-prior, scale-prior,
slab-prior, and inverse-gamma entropy terms:
\begin{equation}
\elbo_{\mathrm{rhs\_ns}}
=
\elbo_{\beta,\mathrm{hs}}
+\elbo_{\beta,\mathrm{slab}}
+\elbo_{\lambda\mid\nu}
+\elbo_{\nu}
+\elbo_{\tau\mid\xi}
+\elbo_{\xi}
+\elbo_{\zeta}
+\elbo_{\mathrm{ent}}
+\elbo_{\beta_0}.
\label{eq:rhs_ns_elbo}
\end{equation}
All terms are available in closed form because each scale factor is
inverse-gamma. This block is added to the static AL or exAL ELBO, whose
likelihood and scale--skewness components are handled by the surrounding
static fitting routine. The fitted static diagnostic object uses the resulting
posterior coefficient draws to produce the coefficient-level summaries shown
in Example 4.

\section{Forecasting, diagnostics, and posterior-predictive synthesis}
\label{sec:forecast_diagnostics_synthesis}

\subsection{Forecasting}

At forecast origin $\tilde t$, the dynamic state forecast recursion gives
\[
\vect{\theta}_{\tilde t+k}^{(m)}
\sim
\N\{a_{\tilde t}^{(m)}(k),R_{\tilde t}^{(m)}(k)\},
\qquad k=1,\ldots,K,
\]
for each posterior or variational draw $m$. Posterior predictive draws are
then generated from
\begin{equation}
Y_{\tilde t+k}^{\mathrm{rep},(m)}
\sim
\exAL_{p_0}\{
\mat{F}_{\tilde t+k}^\top\vect{\theta}_{\tilde t+k}^{(m)},
\sigma^{(m)},\gamma^{(m)}
\}.
\label{eq:forecast_postpred}
\end{equation}
For the AL/DQLM special case, $\gamma^{(m)}=0$. The function
\code{exdqlmForecast()} implements this recursion, and the standard
\code{predict()} method for dynamic fitted objects calls the same forecast
constructor.

\subsection{Diagnostics}

The fitted dynamic diagnostic sequence is based on maximum a posteriori (MAP)
one-step-ahead standardized forecast errors. If $\widehat f_t$ and
$\widehat q_t$ are the MAP one-step-ahead location and variance, then
\begin{equation}
\widehat e_t = \frac{y_t-\widehat f_t}{\sqrt{\widehat q_t}},
\qquad
\widehat u_t = \Phi(\widehat e_t).
\label{eq:pit_plugin}
\end{equation}
The diagnostic plot method uses $\{\widehat e_t\}$ and $\{\widehat u_t\}$ for
the QQ plot, autocorrelation function plot, and standardized forecast-error
plot. The reported Kullback--Leibler (KL) values are deterministic
one-dimensional calibration diagnostics comparing the empirical distribution
of $\{\widehat e_t\}$ with $N(0,1)$; the package uses a deterministic normal
quantile grid rather than a random reference sample.

For posterior predictive draws $y_t^{(1)},\ldots,y_t^{(M)}$, let
$\widehat q_t(\tau_k)$ be the empirical posterior predictive quantile at
level $\tau_k$. The package approximates the continuous ranked probability
score (CRPS) by the integrated quantile-score identity
\begin{equation}
\widehat{\CRPS}_t
=
2\sum_{k=1}^{K}
w_k
\rho_{\tau_k}\{y_t^{\mathrm{obs}}-\widehat q_t(\tau_k)\},
\label{eq:crps_estimator}
\end{equation}
where the weights sum to one \citep{gneiting2007,laio2007verification}. The
posterior predictive loss criterion (PPLC) at the fitted quantile level is
\begin{equation}
\widehat{\PPLC}
=
\sum_t
\frac{1}{M}\sum_{m=1}^M
\rho_{p_0}\{y_t^{\mathrm{obs}}-y_t^{\mathrm{rep},(m)}\}.
\label{eq:pplc_estimator}
\end{equation}
The held-out forecast diagnostic function applies the same CRPS calculation
to forecast objects returned with posterior predictive draws.

\subsection{Posterior-predictive synthesis}

The function \code{quantileSynthesis()} combines posterior predictive draws
from separately fitted models at ordered levels $p_1<\cdots<p_K$. The result
is a post hoc predictive distribution, not a joint multi-quantile posterior.

\begin{appalgorithm}{Posterior-predictive synthesis}
\label{alg:predictive-synthesis}
\par\noindent\emph{Input:} posterior predictive draw matrices or fitted
forecast objects from models at ordered quantile levels
$p_1<\cdots<p_K$.
\par\noindent\emph{Output:} synthesized posterior predictive draws and
summary intervals on a common predictive scale.
\begin{enumerate}[label=(\arabic*)]
\item Extract posterior predictive draw matrices for each fitted quantile.
\item For each time point and draw index, form the quantile-specific
predictive curve over $p_1,\ldots,p_K$.
\item Apply the requested monotonicity correction when fitted predictive
quantiles cross.
\item Interpolate across quantile levels to obtain draws from one predictive
distribution.
\item Return synthesized posterior predictive draws and summary intervals.
\end{enumerate}
\end{appalgorithm}

\section{Laplace--delta and backend notes}
\label{sec:numerical_blocks}

Let $z$ denote the transformed nonconjugate parameter vector with mode
$\hat z$ and negative Hessian $\mat{H}$ at the mode. The LDVB block uses
\begin{equation}
q(z)=\N(\hat z,\mat{H}^{-1}).
\label{eq:ld_gaussian}
\end{equation}
For a smooth scalar function $h(z)$,
\begin{equation}
\E\{h(z)\}
\approx
h(\hat z)
+\frac{1}{2}\tr\{\nabla^2h(\hat z)\mat{H}^{-1}\}.
\label{eq:delta_generic}
\end{equation}
The entropy contribution for the transformed Gaussian block is
\begin{equation}
H\{q(z)\}
=
\frac{1}{2}\log\{(2\pi e)^d|\mat{H}^{-1}|\}.
\label{eq:entropy_ld}
\end{equation}
The Jacobian terms for transforming from $(\ell_\sigma,z_\gamma)$ back to
$(\sigma,\gamma)$ are included in the objective in
\eqref{eq:dyn_exdqlm_ld_objective}. Other ELBO terms use standard entropy
identities for Gaussian, inverse-gamma, GIG, and positive truncated-normal
factors under the parameterizations in
Section~\ref{sec:app_distributional_conventions}. Those bookkeeping terms are
implemented in the package and exercised by the package tests rather than
rederived here.

\label{sec:slice}

The MCMC routines use bounded one-dimensional slice sampling for selected
nonconjugate skewness updates. This is a standard stepping-out/shrinkage
slice sampler restricted to the admissible interval $(L,U)$ for $\gamma$; the
target kernels are the conditional log posteriors induced by the exAL
augmentation.

\label{sec:app_backend_options}
The package also exposes backend options for C++ Kalman recursions, builders,
latent-variable samplers, posterior predictive simulation, and MCMC routing.
These include the \code{exdqlm.use\_cpp\_*} option family, such as
\code{exdqlm.use\_cpp\_kf}, and the \code{exdqlm.cpp\_threads} thread cap.
They are documented in the package help pages and README because they are
runtime controls rather than additional model assumptions.

\bibliography{references}

\end{document}